\documentclass[11pt,twoside]{article}
\usepackage{cozumel2005}
\usepackage{epsf}
\usepackage{psfig}
\usepackage{lscape}

\newcommand{\dvbump}{$\Delta{V}^{\mathrm{bump}}_{\mathrm{HB}}$}

\begin{document}
\title{Uncertainties and systematics in stellar evolution models}   
\author{S. Cassisi}   
\affil{INAF - Astronomical Observatory of Collurania, \\
Via M. Maggini, 64100 Teramo, Italy, cassisi@te.astro.it}    

\begin{abstract} 

In this last decade, our knowledge of evolutionary and structural properties of stars of different
mass and chemical composition is significantly improved. This result has been achieved as a consequence of our 
improved capability in understanding and describing the physical behavior of stellar matter in the different
thermal regimes characteristic of the different stellar mass ranges and/or evolutionary stages.

This notwithstanding, current generation of stellar models is still affected by significant and, usually, not negligible
uncertainties. These uncertainties are related to our poor knowledge of some physical procceses occurring in the real stars
such as, for instance, some thermodynamical processes, nuclear reaction rates, as well as the efficiency of mixing processes.
These drawbacks of stellar models have to be properly taken into account when comparing theory with observations in order to derive
relevant information about the properties of both resolved and unresolved stellar populations.

In this paper we review current uncertainties affecting low-mass stellar models, i.e. those structures with mass in the range between
$0.6M_\odot$ and $\sim1.4M_\odot$. We show what are the main sources of uncertainty along the main evolutionary stages, and discuss the
present level of agreement between theory and observations concerning some selected features of the Color-Magnitude Diagram of
low-mass stars.

\end{abstract}

\section{Introduction} 

During the second half of last century, stellar evolution theory has allowed us to 
understand the Color Magnitude Diagram (CMD)
of both galactic globular clusters (GGCs) and open clusters, so that now 
we can explain the distribution of stars in the observed CMDs in terms of the nuclear evolution of
stellar structures and, thus, in terms of cluster age and chemical composition.
In recent years, however, the impressive improvements achieved for both photometric
and spectroscopic observations, has allowed us to collect
data of an unprecedent accuracy, which provide at the same time a stringent test and a challenge 
for the accuracy of the models.

On the theoretical side, significant improvements have been achieved
in the determination of the Equation of State (EOS) of the stellar
matter, opacities, nuclear cross sections, neutrino emission rates,
that are, the physical inputs needed in order to solve the equations of
stellar structure. At the same time, models computed with this updated physics have
been extensively tested against the latest observations. 

The capability of current stellar models to account for
all the evolutionary phases observed in stellar clusters is undoubtedly 
an exciting achievement  which crowns with success the development of 
stellar evolutionary theories as pursued all along the second half of the last
century. Following such a success, one is often tempted to use
evolutionary results in an uncritical way, i.e., taking these results at
their face values without accounting for theoretical uncertainties. 
However, theoretical uncertainties do exist, as it is clearly shown by the
not negligible differences still existing among evolutionary results 
provided by different theoretical groups. 

The discussion of these theoretical uncertainties was early addressed by
Chaboyer (1995) in a pioneering paper investigating the reliability of  
theoretical predictions concerning H-burning structures presently evolving 
in GGCs (i.e. low-mass, metal-poor stars) and, in turn, on the accuracy of 
current predictions about GC ages. More recently, such an investigation 
has been extended to later phases of stellar evolution by Cassisi et al. 
(1998, 1999), and Castellani \& Degl'Innocenti (1999). Recently, Cassisi (2004) 
has reviewed the issue of the main
uncertainties affecting the evolutionary properties of intermediate-mass stars.

\begin{figure}[!ht]
\plotone{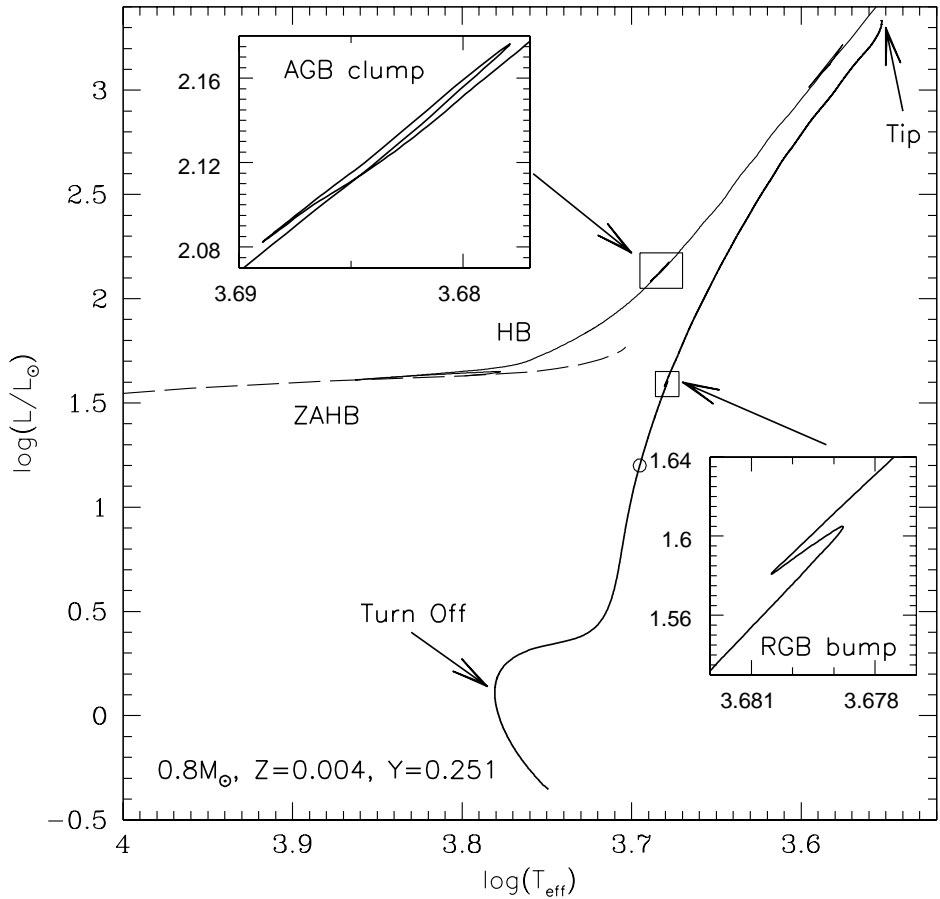}
\caption{The evolution in the H-R diagram of a $0.8M_\odot$ stellar model
during the core and shell H- and He-burning phases. The two insets show the behaviour of the
evolutionary track during the Red Giant Branch bump and the Asymptotic Giant Branch clump (see text for more details).}
\label{fig1}
\end{figure}

In the next sections, we will discuss in some detail the main \rq{ingredients}\rq\ necessary for
computing stellar models and show how the residual uncertainties on these inputs affect
theoretical predictions of the evolutionary properties of low-mass stars.
In particular we will devote a significant attention to the analysis of the
evolutionary phases corresponding to the core H-burning stage with special emphasis on
the late phases of this burning process, to the shell H-burning and to both the central and
shell He-burning stages (see fig.~\ref{fig1}). For the various evolutionary phases, we will discuss
what are the main inputs, adopted in the evolutionary computations, which have the largest
impact on the theoretical predictions.

\section{Stellar evolution: the ingredients} 

The mathematical equations describing the physical behaviour of any stellar structure
are well known since long time, and a clear description of the physical meaning of each one 
of them can be found in several books (as, for instance, Cox \& Giuli 1968 and 
Kippenhan \& Weigart 1990).

The (accurate) numerical solution of these differential equations is no more a problem and 
it can be easily and quickly achieved when using modern numerical solution schemes and current
generation of powerful computers. So, from the point of view of introducing a certain amount
of uncertainty in the computations of stellar models, the solution of the differential equations
constraining the stellar structure is not a real concern.


This notwithstanding, in order to solve these equations, boundary conditions have to be provided:
the boundary condition at the stellar centre are trivial (see the discussion in Kippenhan \& Weigart
1990 and Salaris \& Cassisi 2005); however the same does not apply for those at the stellar surface.
Let us briefly remember that \lq{to provide the outer boundary conditions}\rq\ means to provide the
values of temperature and pressure at the base of the stellar atmosphere: this requirement can be
accomplished either by adopting an empirical relation for the thermal stratification like that provided
by Krishna-Swamy (1966) or a theoretical approximation as the so-called Eddington approximation.

A more rigorous procedure is to use results from model atmosphere computations to obtain the outer 
boundary conditions (see Morel et al. 1994). 
In general model atmospheres are computed considering a plane-parallel
geometry, and solving  the hydrostatic equilibrium equation together with the frequency dependent 
(no diffusion approximation is allowed in these low-density layers) equation of radiative
transport and convective transport when necessary, plus the appropriate equation of state.

In the following, we will discuss the impact of different choices about the outer boundary conditions
on various theoretical predictions.

\subsection{The physical inputs} 

In order to compute a stellar structure, it is fundamental to have an accurate description of the
physical behaviour of the matter in the thermal conditions characteristics of the stellar interiors
and atmospheres. This means that we need to know as much accurately as possible several physical
inputs as:

\begin{itemize}

\item{{\bf the opacity}: the \lq{radiative}\rq\ opacity is related to the mean free path of photons
inside the stars and it plays a pivotal role in determining the efficiency of heat transfer via
radiative processes. When the stellar matter is under conditions of partial or full electron
degeneracy, electrons are able to transport energy with a large efficiency since they have a
longer mean free path than in case of non degenerate electrons. In this case, the energy transport
by conduction becomes quite important and the value of the conductive opacity has to be properly
evaluated.}

\item{{\bf the equation of state}: the EOS of the stellar matter is another key input for 
the model computations; it connects  pressure, density, temperature and chemical composition 
at each point within the star, determines the value of the adiabatic gradient 
(which is the temperature gradient in most of the
convective region), the value of the specific heat (which appears in the   
expression of the gravitational energy term), and plays a crucial role in the evaluation of    
the extension of the convective regions.}

\item{{\bf the nuclear cross-sections}: the  evaluation of the cross-sections for the various
nuclear burning processes occurring in the stellar interiors is quite important in order to
properly establish the energy balance in the star. Thanks to laboratory experiments, many nuclear
cross-sections are nowadays known with a high accuracy. However, there are still some important
nuclear processes for both the H- and the He-burning, whose nuclear rate is poorly known.}

\item{{\bf the neutrino energy losses}: a precise determination of the energy losses due to 
neutrino emission is also important when the star is characterized by high density and low
temperature as it occurs in the interiors of Red Giant stars.}

\end{itemize}

It exists a quite rich literature describing the improvements which have been achieved in this
last decade concerning our knowledge of the physical inputs required for computing stellar models.
Therefore, in the following, unless quite relevant for our discussion, we will not discuss in detail this issue and
refer the interested reader to the exhaustive reference lists reported by Chaboyer (1995), Catelan et al. (1996), 
Cassisi et al. (1998, 1999, 2001), Salaris, Cassisi \& Weiss (2002) and references therein.

\subsection{The microscopic mechanisms}

When computing a stellar model, some important assumptions have to be done concerning the
efficiency of some microscopic mechanisms. With the expression \lq{microscopic mechanisms}\rq\ 
we refer to all those mechanisms which, working selectively on the different chemical species, can
modify the chemical stratification in the stellar interiors and/or atmosphere. These mechanisms
are: atomic diffusion and radiative levitation.

\begin{itemize}

\item{{\bf atomic diffusion}: atomic diffusion is a basic physical
transport mechanism driven by collisions of
gas particles. Pressure, temperature and chemical abundance gradients are the driving
forces behind atomic diffusion. A pressure gradient and a temperature
gradient tend to push the heavier elements in the direction of
increasing pressure and increasing temperature, whereas the resulting
concentration gradient tends to oppose the above processes. The speed
of the diffusive flow depends on the collisions with the surrounding
particles. The efficiency of the different mechanisms involved in the atomic diffusion
process is given in terms of atomic diffusion coefficients which have to be estimated
on the basis of laboratory measurements.}

\item{{\bf radiative levitation}: it is an additional transport mechanism caused
by the interaction of photons with the gas particles, which acts
selectively on different atoms and ions. Since
within the star a net energy flux is directed towards the
surface, photons provide an upward \lq{push}\rq\ to the gas
particle with which they interact, effectively reducing the
gravitational acceleration. Since, at the basis of this process there are the interactions
of photons with gas particles, it is clear that the efficiency of
radiative levitation is related to the opacity of the stellar
matter, in particular to the monochromatic opacity, and increases
for increasing temperature (let us remember that radiation pressure $P_{rad}\propto{T^4}$). The
evaluation of the radiative accelerations is really a thorny problems due to the need of accounting
for different interaction processes between photons and chemical elements and how the momentum of
photons is distributed among ions and free electrons.}

\end{itemize}

Until few years ago, all these non-canonical processes were usually ignored in stellar models
computations. However, helioseismology has clearly shown how important is to include atomic
diffusion in the computation of the so-called Standard Solar Model (SSM), in order to obtain a good
agreement between the observed and the predicted frequencies of the non-radial p-modes (see for
instance Christensen-Dalsgaard, Proffitt \& Thompson 1993). In the meantime, quite recent spectroscopical
measurements of the iron content in low-mass, metal-poor stars in galactic globular clusters
strongly point out the importance of including radiative levitation in stellar computations in
order to put in better agreement empirical estimates with the predictions provided by diffusive
models.

\subsection{The macroscopic mechanisms}

When computing a stellar model, one has unavoidably to account for the occurrence of mixing in the
real stars. Due to current poor knowledge of how to manage the mixing processes in a stellar
evolutionary code, the efficiency of convection is commonly treated by adopting some approximate
theory. In this context, it has to be noticed that when treating a region where convection is stable,
one has to face with two problems:

\begin{itemize}

\item What is the \lq{right}\rq\ temperature gradient in such region?

\item What is the \lq{real}\rq\ extension of the convective region?

\end{itemize}

The first question is really important only when considering the outer convective regions such as
the convective envelopes of cool stars. This occurrence is due to the evidence that, in the stellar
interiors as a consequence of the high densities and, in turn, of the high capability of convective
bubbles to transport energy, the \lq{real}\rq\ temperature gradient has to be equal to the
adiabatic one. This consideration does not apply when considering the outer, low-density, stellar
regions, where the correct temperature gradient has to be larger than the adiabatic one: the
so-called {\sl superadiabatic gradient}. One of the main problem we still have in computing star
models is related to the correct estimate of this superadiabatic gradient. It is important to
notice that this is not an academic question since the radius and, in turn, the effective
temperature of cool stars (let us say: stars with $T_{eff}<8000K$) is drastically affected by the
choice of the superadiabatic gradient.

Almost all evolutionary computations available in literature rely on the mixing length theory 
(MLT; B\"ohm-Vitense 1958). It contains a number of free parameters, whose   
numerical values affect the model $T_{\rm eff}$; one of them  
is $\alpha_{\rm MLT}$, the ratio of the mixing length to the pressure scale   
height, which provides the scale length of the convective motions  
(increasing $\alpha_{\rm MLT}$ increases the model $T_{\rm eff}$).   
There exist different versions of the MLT, each one assuming different   
values for these parameters; however, as demonstrated by Pedersen,   
Vandenberg \& Irwin~(1990), the $T_{\rm eff}$ values obtained from the   
different formalisms can be made consistent, provided that a suitable value of   
$\alpha_{\rm MLT}$ is selected. Therefore, at least for the evaluation of    
$T_{\rm eff}$, the MLT is basically a one-parameter theory.   
   
The value of $\alpha_{\rm MLT}$  is usually calibrated by reproducing the solar $T_{\rm eff}$, 
and this solar-calibrated value is then used for computing models 
of stars very  different from the Sun (e.g. metal poor Red Giant Branch (RGB) and Main Sequence (MS) stars of   
various masses). We will come back on this issue in the following.

It is worth recalling that there exists also an alternative formalism for the computation 
of the superadiabatic   
gradient, which in principle does not require the calibration of any free   
parameter. It is the so-called Full-Spectrum-Turbulence theory (FST,   
see, e.g., Canuto \& Mazzitelli~1991, Canuto, Goldman \&   
Mazzitelli~1996), a MLT-like formalism with a more sophisticated   
expression for the convective flux, and the scale-length   
of the convective motion fixed a priori (at each point in a convective region,   
it is equal to the harmonic average between the distances from the top and the bottom convective   
boundaries). From a practical point of view, the FST theory contains also a free parameter
which has to be fixed, even if it seems to have a physical meaning larger than that
of $\alpha_{\rm MLT}$.

The problem of the real extension of a convective region really affects both convective core and
envelope. In the canonical framework it is assumed that the border of a convective region is fixed
by the condition - according to the classical Schwarzschild criterion - that the radiative gradient
is equal to the adiabatic one. However, it is clear that this condition marks the point where the
acceleration of the convective cells is equal to zero, so it is realistic to predict that the
convective elements can move beyond, entering and, in turn, mixing the region surrounding the
classical convective boundary. This process is commonly referred to as convective overshoot.

Convective core overshoot is not at all a problem for low-mass stars during the H-burning phase
since the burning  process occurs in a radiative region. However, the approach used for treating
convection at the border of the classical convective core is important during the following
core He-burning phase as discussed in the next sections. Convective envelope overshoot could be
important for low-mass stars, since these structures have large convective envelope during the
shell H-burning phase and the brightness of the bump along the RGB could be significantly affected
by envelope overshoot (this topic will be addressed in more detail in section~\ref{bump}).

In low-mass stars, during the core He-burning phase, the occurrence of convection-induced mixing is
a relevant problem when computing stellar models along this evolutionary stage. We discuss in more 
detail this issue in section~\ref{zahb}.

Near the end of the core He-burning phase, there is another process associated with
mixing, that could potentially largely affect the evolutionary properties of the models: a sort of
pulsating instability of convection, the so-called {\sl breathing pulse} (Castellani et al. 1985), can occur, 
driving fresh helium  into the core and so,
affecting the core He-burning lifetime as well as the carbon/oxygen ratio at the center of the 
star. It is still under debate if this mixing instability occurs in real stars or, if it is a
fictitious process occurring as a consequence of our poor treatment of convection, as for instance,
of the commonly adopted assumption of instantaneous mixing.

\section{H-burning structures: the Turn-Off}
\label{msto}

The brightness of the bluest point along the MS, the so-called {\sl Turn Off} (TO) (see fig.~\ref{fig1}),
is the most important clock marking the age of the stellar clusters. It is well known that in
order to use this observational feature of the Color-Magnitude diagram for estimating the cluster
age, one needs to know the distance to the cluster and the chemical composition of the stars
belonging to the stellar system. The impact of current uncertainties in both stellar 
cluster distances and chemical composition measurements on the age estimates has been extensively
discussed in literature so it will not be repeated here and we refer to the interesting work by
Renzini (1991). So, from now on, we will assume to know \lq{perfectly}\rq\ both the distance and
the chemical composition of the stellar clusters and will concentrate our discussion on the
reliability and accuracy of the age - luminosity (of the TO) calibration provided by evolutionary
stellar models.

It is clear that the check of the accuracy of the evolutionary models should correspondingly
become a cornerstone in our attempt of obtaining accurate ages for globular clusters as well as
robust results concerning the star formation history of composite stellar populations.

The main \rq{ingredients}\rq\ adopted in stellar models computations which affect the
observational properties of stellar models at the TO and, in turn, the age -
luminosity calibration are the following :

\begin{itemize}

\item{EOS $\longrightarrow$ luminosity, effective temperature}

\item{Radiative opacity $\longrightarrow$ luminosity, effective temperature}

\item{Nuclear reaction rates $\longrightarrow$ luminosity}

\item{Superadiabatic convection $\longrightarrow$ effective temperature}

\item{Chemical abundances $\longrightarrow$ luminosity, effective temperature}

\item{Atomic diffusion $\longrightarrow$ luminosity, effective temperature}

\item{Treatment of the boundary conditions $\longrightarrow$ effective temperature}

\end{itemize}

For each ingredient, we have also listed the observational property of the TO structure which is
affected by a change of the corresponding ingredient. Therefore, it appears evident that some
\lq{inputs}\rq\ affect directly the age - luminosity relation because they modify the bolometric
magnitude of the TO for a fixed age; some other inputs really can modify also (or only) the effective
temperature of the TO models, so they affect the age - luminosity relation indirectly through the
change induced in the bolometric correction adopted for transferring the theoretical predictions
from the H-R diagram to the various observational planes.
Now we discuss in some details the effect of current uncertainties in these inputs in the
calibration of the age - luminosity relation.

\subsubsection{The EOS:} the importance of an accurate EOS when computing SSM has been
largely emphasized by all helioseismological analysis (see for instance Degl'Inno\-cen\-ti et al. 1997 and references
therein). However, Chaboyer \& Kim (1995) were the first to strongly point out the relevance of an
accurate EOS for computing H-burning stellar models of low-mass, metal-poor stars due to the huge
impact on the age - luminosity calibration and, in turn, on the dating of GGCs. 
More in detail, they have clearly shown how the proper treatment of non-ideal effects
such as Coulomb interactions significantly affects the thermal properties of low-mass stars and
then their core H-burning lifetime.

Chaboyer \& Kim (1995) showed how the use of the OPAL EOS (Rogers 1994) - the most
updated EOS available at that time - would imply a reduction of the GC age of about 1Gyr (i.e. of
about 7\% when compared with the ages derived by using models based on less accurate EOSs).

The OPAL EOS has been largely updated along this decade (see Rogers \& Nayfonov 2002). However,
the results for H-burning structures do not change significantly with respect the predictions
obtained at the time of the first EOS release. It is worth emphasizing that the OPAL EOS results
have been recently confirmed by independent analysis such that performed by A. Irwin (Irwin 2005,
Cassisi, Salaris \& Irwin 2003). One can also notice that almost all more recent library of
stellar models are based on updated EOS. Therefore, we think that current residual uncertainties
on the EOS for low-mass stars do not significantly affect the reliability of current age -
luminosity calibrations.

\subsubsection{The radiative opacity:}
it is one of the most essential ingredients of the model input physics. As a general rule,
increasing the radiative opacity makes dimmer stars (roughly speaking $L\propto{1/\kappa}$), 
which then take longer to burn their central
hydrogen. So for a given stellar mass, the TO luminosity is decreased, and the time needed to
reach it is increased. So the two effects tend to balance each other, and the age - luminosity
calibration is less affected. However, larger opacities favor the envelope expansion, and
therefore the MS TO is anticipated. In this last decade, a big effort has been devoted to a better
determination of both high- and low-temperature ($T<10000K$) opacities. Concerning
the high-temperature opacity the largest contribution has been provided by the OPAL group 
(Iglesias \& Rogers 1996) whose evaluations represent a sizeable improvement with respect the classical
Los Alamos opacity. So the question is: how much accurate are current evaluations of radiative
opacity in the high-T regime?

Recently this issue has been investigated by two independent analysis for thermal conditions and
chemical compositions appropriate in the Sun: Rose (2001) considered several opacity tabulations
and found that for temperatures typical of the solar core there is a standard deviation of about
5\% around the average; Neuforge-Verheecke et al. (2001) performed an accurate comparison between
the OPAL opacity and that provided by Magee et al. (1995) and disclosed that the mean difference
between the two opacity set is of about 5\%,  being the OPAL opacity larger than the Magee et
al.'s on almost the whole temperature range.

When considering metal-poor stars, due to the lack of heavy elements, one can expect that the
opacity evaluation is simpler than for metal-rich stars and, then the estimates should be more
robust. In fact, as verified by Chaboyer \& Krauss (2002) the difference between the OPAL and the
LEDCOP opacities in the metal-poor regime ranges from $\approx1$\% at the star centre to about 4\%
at the base of the convective envelope.

However, the existence of a good agreement between independent estimates does not represent an
evidence that the predicted opacity is equal to the \lq{true}\rq\ one: there is a general
consensus that, at least, for conditions appropriate for the core of metal-poor stars, current
uncertainty should not be larger than about 5\%. For temperatures of the order of $10^6K$, a
larger uncertainty seems to be possible: quite recently Seaton \& Badnell (2004) have shown that, 
for temperature of this order of magnitude, a difference of the order pf $\sim13$\% does exist between 
the monochromatic opacities provided by the OPAL group with respect those provided by the Opacity Project.

As verified by Chaboyer \& Krauss (2002) on the basis of an extended set of Monte Carlo
simulations, an increase of about 2\% in the high-T opacity would imply an increase of about 3\%
in the age determination based on a theoretical calibration of the relation between the cluster
age and the mass of the star evolving along the Sub-giant. Since the sensitivity of the TO
brightness to change in the adopted radiative opacity is lower than the age indicator considered
by Chaboyer \& Krauss (2002), we expect that a change of 2\% in the high-T opacity leaves almost
unaffected the age - luminosity calibration (see also Chaboyer 1995 and Cassisi et al. 1998).

As far as it concerns low-temperature opacities, since they affect mostly cool stars like RGB ones
 - TO stars have effective temperatures large enough for not being significantly affected by
different choices about low-T opacities - we postpone a discussion about the impact of current 
uncertainty on stellar models, to the section devoted to RGB stellar models. Here it is suffice to
note that current errors on this ingredient have a negligible effect (of the order of 1\%) on age
estimates as verified by Chaboyer \& Kim (1995).

\subsubsection{Nuclear reaction rates:}

the reliability of theoretical predictions about evolutionary lifetimes
critically depends on the accuracy of the nuclear reaction rates since nuclear burning provides the
bulk of the stellar luminosity during the main evolutionary phases. In these last years, a
large effort has been devoted to increase the measurement accuracy at energies as close as
possible to the Gamow peak, i.e. at the energies at which the nuclear reactions occur in the
stars. 

The effect on the age - luminosity calibration of current uncertainties on the rates of the
nuclear reactions involved in the {\sl p-p} chain has been extensively investigated by several
authors (Chaboyer 1995, Chaboyer et al. 1998, Brocato, Castellani \& Villante 1998). The main
result was that, for a realistic estimate of the possible errors on these rates, the effect on the
derived ages was almost negligible (lower than $\sim2$\%). The explanation of this result is
simply that the nuclear processes involved in the {\sl p-p} chain are really well understood so
the associated uncertainty is quite small.

However, near the end of core H-burning stage, due to the lack of H, the energy supplied by the
H-burning becomes insufficient and the star reacts contracting its core in order to produce the
requested energy via gravitation. As a consequence, both the central temperature and density
increase and, when the temperature attains a value of the order of $13-15\times10^6K$, the
H-burning process is really governed by the more efficient {\sl CNO} cycle, whose efficiency is
critically depending on the reaction rate for the nuclear process $^{14}N(p,\gamma)^{15}O$, since
this is the slowest reaction in the {\sl CNO} cycle. 

The TO luminosity depends on the rate of this nuclear process: the larger the rate, the fainter
the TO is (roughly speaking $\Delta\log(L_{TO}/L_\odot)\propto - 0.015\delta{CNO}$ (Brocato et al.
1998)). On the contrary, the core H-burning lifetime is marginally affected by the rate of this
process, being mainly controlled by the efficiency of the {\sl p-p} chain. For an exhaustive discussion
on this issue, we refer to Weiss et al. (2005). 

Until a couple of year ago, the rate for the $^{14}N(p,\gamma)^{15}O$ reaction was uncertain, at
least, at the level of a factor of 5. In fact, all available laboratory measurements were
performed at energies well above the range of interest for astrophysical purpose and, therefore,
a crude extrapolation was required (Caughlan \& Fowler 1988, Angulo et al. 1999). Due to the presence of a complex resonance in the nuclear
cross section at the relevant low energies, this extrapolation was really unsafe (see Angulo et al. 1999,
while for a discussion
of the impact of the estimated uncertainty on the age-luminosity relation we address the reader to
the paper by Chaboyer et al. 1998).

Luckily enough, recently the LUNA experiment (Formicola et al. 2003) has significantly improved
the low energy measurements of this reaction rate, obtaining an estimate which is about a factor
of 2 lower than previous determinations. The effect on H-burning stellar models and, in turn, on the age -
luminosity relation has been investigated by Imbriani et al. (2004) and Weiss et al. (2005).

The lower rate for the $^{14}N(p,\gamma)^{15}O$ reaction leads to a brighter and hotter TO for a fixed
age. The impact of this new rate on the age - luminosity relation is the following: for a
fixed TO brightness the new calibration predicts systematically older cluster ages, being the difference
with respect the \lq{old}\rq\ calibration of the order of 0.8-0.9Gyr on average.

\subsubsection{Superadiabatic convection:}

as already stated, the convection in the outer layers is commonly managed by adopting the mixing
length formalism in which a free parameter is present: the mixing length. Its value is usually calibrated
on the Sun\footnote{Since atomic diffusion (see the following discussion) modifies the envelope chemical stratification and,
in turn, the envelope opacity, the value obtained for the solar calibrated mixing length does depend on 
if atomic diffusion is taken into account when computing the SSM. Needless to say that the most correct approach
is to calibrate the mixing length on a diffusive SSM.}
(see for instance, Salaris \& Cassisi 1996, and Pietrinferni et al. 2004). However, since there
is no compelling reason according to which the mixing length should be the same for the Sun  and metal-poor
stars or constant for different evolutionary phases, it is worthwhile to investigate the impact of
different choice about the mixing length calibration (see also below). 

One has to bear in mind that a change in the mixing length, i.e. a change in the superadiabatic convection
efficiency, alters only the stellar radius and, in turn, the effective temperature, leaving unchanged the
surface luminosity. This is shown in fig.~\ref{fig2}, where we plot two isochrones computed adopting two
different values for the mixing length.

Since, the effect on the stellar radius due to a mixing length variation depends on the extension of the
superadiabatic region - that is larger for stars in the mass range: $\sim0.7M_\odot - \sim1.4M_\odot$ (less
massive stars being more dense objects are almost completely adiabatic, while in more massive stars the
superadiabatic region is extremely thin) and, in turn, on the total star mass, any change in the efficiency
of outer convection alters the shape of the theoretical isochrones in the region around the TO (see
fig.~\ref{fig2}).

{\small
\begin{figure}
\hspace*{1.5cm}\psfig{figure=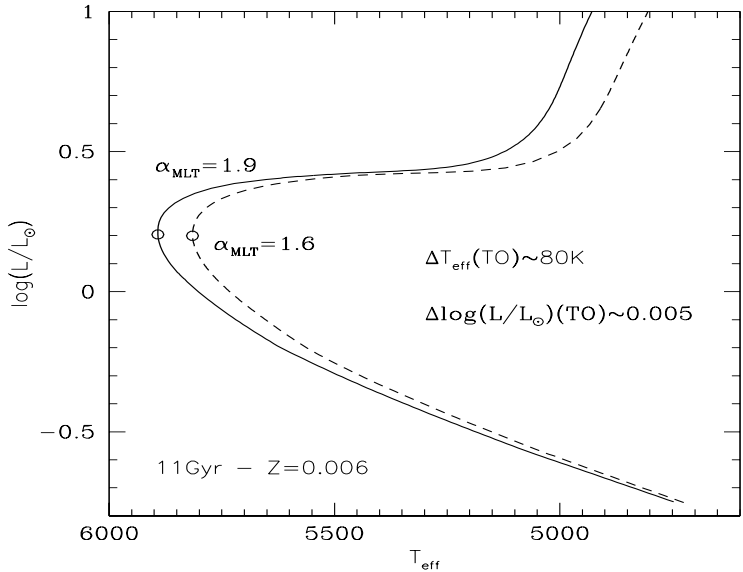,height=8truecm,width=10truecm}
\vspace*{-0.5cm}\caption{The morphology of two isochrones corresponding to the same cluster age
(see label), computed by assuming two different values of the mixing length.}
\label{fig2}
\end{figure}
}

Therefore, from the point of view of the age - luminosity relation, the uncertainty in the superadiabatic
convection efficiency introduces a certain amount of indetermination as a consequence of the induced change
in the $T_{eff}$ and, then, in the adopted bolometric correction.
According to Chaboyer (1995), the maximum uncertainty related to the treatment of convection in stellar
models is of the order of 10\%. However, one has to note that this estimate was obtained by changing the
mixing length value in the range from 1 to 3.
Really, a so huge variation of the mixing length seems not to be requested by current physical framework:
almost all independent set of stellar models have been computed by using similar mixing length values (the spread is of
the order of 0.2-0.3); in addition within a given theoretical framework there is no need to change the
mixing length of almost a factor of 2 (see for instance the analysis of the mixing length calibration
performed by Salaris \& Cassisi~1996, in a wide metallicity range).

This issue has been recently revised by Chaboyer et al. (1998): they found (see their fig.~5) that a change
of 0.1 in the mixing length causes a variation of about 1\% in the globular cluster age; since they assume
a realistic uncertainty of about 0.25 in the convection efficiency, this translates in an uncertainty of
$\approx3$\% in the cluster age.

\subsubsection{Diffusive processes:}
since, at least, a decade, helioseismological constraints have brought to light the evidence that diffusion
of helium and heavy elements must be at work in the Sun. So, it is immediate to assume that this physical
process is also efficient\footnote{Really, on theoretical grounds, one expects that atomic diffusion is more
efficient in metal-poor MS stars, because in such structures the extension of the convective envelope is
lower than in metal-rich objects, and it is well known that convection drastically reduces the efficiency
of any diffusive process.} in more metal-poor stars like those currently evolving in galactic GCs.

This notwithstanding, the evaluation of the atomic diffusion coefficient is not an easy task as a
consequence of the complex physics one has to manage when analizing the various diffusive mechanisms, and
moreover the range of efficiency allowed by helioseismology is still relatively large. For the Sun,
Fiorentini et al. (1999) estimated an uncertainty of about 30\% in the atomic diffusion coefficient.
Perhaps the uncertainty is also larger for metal-poor stars due to the lack of any asteroseismological
constraint. On this ground, it is not unrealistic to estimate an uncertainty of about a factor of two
in the atomic diffusion efficiency. From an evolutionary point of view, the larger the atomic diffusion
efficiency, the lower the cluster age estimate is (see Castellani et al. 1997 and references therein). 
The impact of this source of uncertainty on stellar models has been
extensively investigated by Castellani \& Degl'Innocenti (1999): for ages of the order of 10Gyr, 
to change the efficiency of diffusion within the quoted range modifies
the TO brightness of about 0.16 mag, which corresponds to a variation of the cluster age of about
$-0.7/+0.5$Gyr; however for larger ages the situation is worst and for an average age of 15Gyr the error is
equal to $-1.7/+1$Gyr. So it appears evident that current uncertainty in the atomic diffusion coefficients
is one of the largest source of error in the theoretical calibration of the age - luminosity relation.

However, we are now faced with an additional and, perhaps, more important problem concerning atomic
diffusion: as already discussed, helioseismology strongly support SSMs accounting for atomic diffusion, but
recent spectroscopical measurements (Castlho et al. 2000, Gratton et al. 2001, Ramirez et al. 2001) of the
iron content are in severe disagreement with the predictions provided by diffusive models: the measured
iron content does not appear to be significantly reduced with respect the abundance estimated for giant
stars in the same cluster as one has to expect as a consequence of diffusion being at work.
So the question is: how to reconcile these two independent evidence? 

This can be achieved only accounting for two important pieces of evidence:

\begin{itemize}

\item{according to Turcotte et al. (1998), radiative acceleration in the Sun can amount to about the 40\%
of gravitational acceleration, and one can expect that its value is larger in more metal-poor, MS stars
(see above);}

\item{there are some evidence according to which a slow mixing process (turbulence?) below the solar
convective envelope could help in explaining better the observed Be and Li abundances (Richard et al. 1996)
and improve the agreement between the predicted sound speed profile and that derived from
helioseismological data (Brun et al.~1999).}

\end{itemize}

Both these evidence, coupled with helioseismological analysis, clearly support the computations of MS
stellar models accounting simultaneously for atomic diffusion, radiative levitation and some sort of
extra-mixing. This new generation of models has been recently provided by Richard et al. (2002) and
Vandenberg et al. (2002): the main outcome is that these models are able to reconcile helioseismology with
the recent spectroscopical measurements of the iron in GGCs. In fact, these models predict that, at odds,
with predictions provided by models accounting only for atomic diffusion, the surface abundance of iron 
(and also of other heavy elements) is depleted with a quite lower efficiency and it can also become
overabundant with respect the initial value as a consequence of radiative levitation which pushes iron from
the interior toward the stellar surface (see fig.~8 in the paper of Richard et al. 2002).

Concerning the age - luminosity calibration, it is worth noticing that it is not significantly affected by
the inclusion of radiative levitation in stellar models computations: models accounting for both atomic
diffusion and radiative levitation lead to a reduction of the order of 10\% in the GGC age at a given TO
brightness, i.e. more or less the same reduction which is obtained when accounting only for a (standard)
efficiency of microscopic diffusion.

\subsubsection{The treatment of boundary conditions:}

the effect of adopting different choices about the outer boundary conditions on the age - luminosity
calibration has been extensively investigated by Chaboyer (1995) and Chaboyer et al. (1998). The main
result was that the calibration is only marginally affected by the adopted outer boundary conditions.

\subsubsection{The chemical abundances:}

the evolutionary properties of stars strongly depend on the initial chemical abundances, i.e. on the
initial He content (Y = abundance by mass of helium) and heavy elements abundance ($Z$ = metallicity =
abundance by mass of all elements heavier than helium; in the spectroscopical notation it is indicated as
$[M/H]$). So, the age - luminosity relation depends on both $Y$ and $[M/H]$: the typical dependences (Renzini
1997) are $\partial\log{t_9}/\partial{Y}\approx0.4$ and $\partial\log{t_9}/\partial{[M/H]}\approx0.1$,
where $t_9$ is the cluster age in billion of years.

The initial He content of the old, metal-poor galactic GCs is well known (see Cassisi, Salaris \& Irwin
2003 and references therein) and it has to be in the range Y=0.23 - 0.25. So, assuming an uncertainty of
about 0.02 in the initial He abundance, the previous relation indicates that this uncertainty gives a
negligible 2\% error in age. The metal content of the best studied clusters is uncertain by perhaps
0.2-0.3 dex  - most of it being systematic, which translates into a $\approx9$\% uncertainty in age.
However, when discussing the uncertainty associated to the adopted metallicity, one has also to pay attention
to the \lq{composition}\rq\ of the metallicity, i.e. to the distribution of the various heavy elements in
the mixture. In particular, there are clear indications that $\alpha-$elements (such as O, Ne, Mg, Si, S,
Ar, Ca and Ti) are enhanced in metal-poor stellar systems with respect to the Sun (i.e. $[\alpha/Fe]>0$).

The effect of an $\alpha-$enhanced mixture on the evolutionary properties of stellar models has been
extensively investigated in literature and we refer to Salaris, Chieffi \& Straniero (1993) and Vandenberg
et al. (2000) and references therein: as a general rule, at a given iron content, an increase of the
$\alpha-$elements abundance makes the evolutionary tracks fainter and cooler. As a consequence, for a fixed
TO brightness, the cluster age decreases when $[\alpha/Fe]$ increases: a more accurate statement is that,
at a fixed TO luminosity, the cluster age is reduced by $\approx$7\% ($\sim1$Gyr) for each 0.3 dex 
increase in the $[\alpha/Fe]$ value (see fig.~4 in Vandenberg, Bolte \& Stetson 1996).
An accurate analysis of the stellar models discloses that almost the 60\% of this variation in the age is
due to the change in the radiative opacity associated with the modification of the heavy elements mixture,
while the remaining difference is provided by the change in the efficiency of the {\sl CNO} cycle related
to the increased abundance of O in the $\alpha-$enhanced mixture.

These considerations clearly suggest that if we would know the exact $\alpha-$en\-han\-ced mixture of a stellar
system we could compute stellar models for that mixture once the appropriate $\alpha-$enhanced radiative
opacities are provided and the burning network is correspondingly updated. On practice, this is almost
impossible: 1) it is not possible to compute extended set of stellar models for any specified heavy
elements distribution, 2) radiative opacity tabulations for any $\alpha-$enhanced distribution are
difficult to be provided (mostly in the low-temperature regime). However, there is the possibility
to overcome this problem. In fact, it has been shown by Salaris et al. (1993) that isochrones for enhanced $\alpha-$element
abundances are well mimicked by those for a scaled-solar mixture, simply by requiring the
total abundance of heavy elements to be the same: this is the so-called \lq{rescaling}\rq\ approximation.
This topic has been recently reanalyzed by Vandenberg et al. (2000, but see also Vandenberg \& Irwin 1997)
which have demonstrated that the rescaling approach is quite reliable for metallicity of the order
of $[Fe/H]\sim-0.8$ ($Z\approx0.002$) or lower. For larger metallicity, it is no longer correct to rely on this
assumption and $\alpha-$enhanced stellar models have to be used when comparing theory with observations.

{\small
\begin{figure}
\vspace*{-0.8cm}  
\hspace*{1.5cm}\psfig{figure=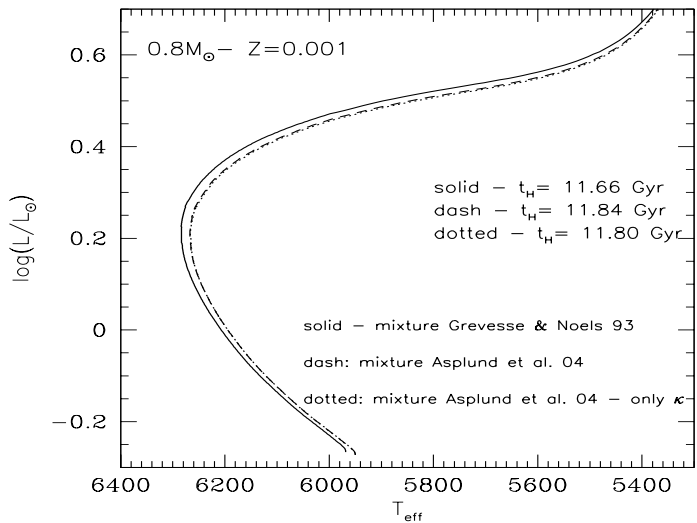,height=8truecm,width=10truecm}
\vspace*{-0.8cm}\caption{ The evolutionary tracks of a $0.8M_\odot$ model with Z=0.001,
computed by adopting alternatively the Grevesse \& Noels (1993) heavy elements mixture or the
Asplund et al.'s (2004) one. The dotted line shows the path of the evolutionary track obtained by accounting for
the Ausplund et al.'s mixture in the radiative opacity but not in the burning network.}
\label{fig3}
\end{figure}
}

Concerning the heavy elements distribution, there is an additional possible source of uncertainty: recent
analyses of spectroscopical data based on 3-D hydrodynamic atmospheric models (Asplund et al. 2004) suggest
that the heavy elements distribution in the Sun is significantly different with respect previous estimates
(Grevesse \& Sauval 1998). More in detail, the abundance of oxygen and of other heavy elements has been
drastically reduced by these new measurements. As a consequence, the metal over hydrogen ratio $(Z/X)_\odot$ has
been significantly changed from $(Z/X)_\odot=0.0230$ to 0.0165, so the Sun's metallicity has been
drastically reduced by a factor of $\sim1.4$. In our belief, these new measurements have to be confirmed by
other independent and accurate analyses\footnote{It is important to note that the new estimates of the solar
metallicity put the SSMs in severe disagreement with helioseismological constraints (Basu \& Antia~2004,
Bahcall, Serenelli \& Pinsonneault~2004). However, a possible solution for this problem has been suggested by
Seaton \& Badnell~2004, through an increase of the radiative opacity at the boundary of the solar convective
envelope.}. However, it is interesting to analyze what is their impact on the
evolutionary scenario.

In fig.~\ref{fig3}, we show the evolutionary track of a $0.8M_\odot$ computed adopting different
assumptions about the distribution of heavy elements in the mixture: the effect of adopting the new Asplund
et al's mixture is quite negligible\footnote{One has to notice that the effect would be slightly larger -
but always very small - if we would consider a more massive stars and/or a higher global metallicity.}; so
one can expect that also the effect on the age - luminosity calibration is irrelevant. This has been
demonstrated via accurate evolutionary computations by Degl'Innocenti et al. (2005).

Although, the effect of the new solar heavy elements distribution on the theoretical age - luminosity
relations is quite negligible, the new estimate of the solar metallicity could potentially have a huge
impact on the GGC age scale: let us assume as a first order approximation that the spectroscopical
measurement of the metallicity of the stellar systems is not affected by the use of these updated set of
3-D model atmospheres (but see the preliminary analysis of Asplund~2004). If so, the value of $[M/H]$ for a cluster remains unchanged. Since the relation
connecting the abundance by mass of heavy elements ($Z$) to the global metallicity in the spectroscopical
notation ($[M/H]$) implies the use of the solar metallicity $Z_\odot$: $Z\sim{Z_\odot}\times10^{[M/H]}$.
Simply due to the change in the value of $Z_\odot$, now when comparing the theoretical framework with the
cluster observations we must use a metallicity $Z_{new}$ that is equal to $\sim0.65Z_{old}$ (being
$Z_{old}$ the metallicity adopted when accounting for the \lq{old}\rq\ solar heavy elements distribution).
This occurrence implies that for a fixed TO brightness, the cluster age is increased of about 0.7Gyr.

{\small
\begin{figure}
\hspace*{1.5cm}\psfig{figure=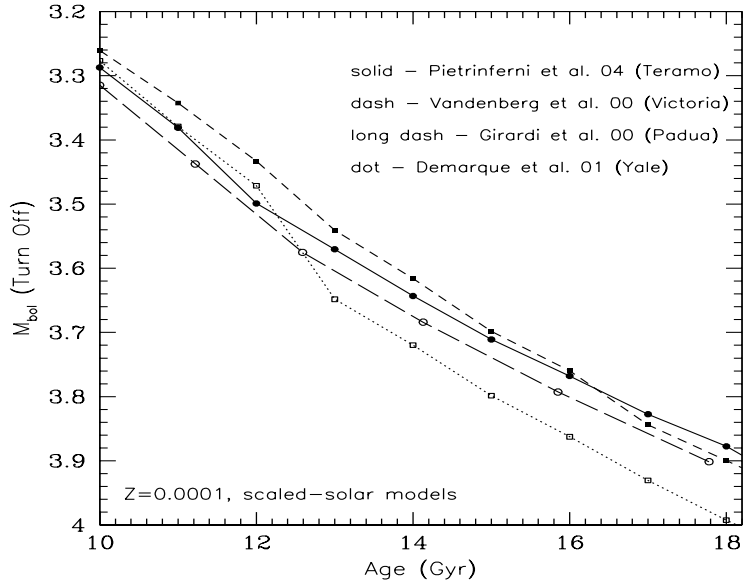,height=8truecm,width=10truecm}
\vspace*{-0.5cm}\caption{The age - Turn Off brightness relation provided by the most updated libraries of stellar
models currently available, for a fixed metallicity (see label).}
\label{fig4}
\end{figure}
}

This notwithstanding, we think that the real problem is another one: if the use of these new generation of
3-D model atmospheres has to drastically modified our knowledge of the solar chemical composition, one
should expect that their use affects also the determination of the metallicity of metal-poor clusters. So,
the main question is: what is - nowadays - the correct metallicity scale for stellar clusters?

\subsection{The age - Turn-Off brightness calibration: the state of art}

In the previous sections, we have discussed the main sources of uncertainty in the theoretical
calibration of the age - TO brightness relation. However, in order to have an idea of the level of
confidence in this important age indicator, we show in fig.~\ref{fig4} the TO brightness - age
calibrations provided by the most updated set of evolutionary models presently available. It is worth
noticing that all the theoretical predictions, but the one provided by the Yale group\footnote{It could be
possible that the mismatch between the Yale results and the other ones is simply due to the evidence that in
the original files where the isochrones are listed, only a few number of lines are reported and this makes a
problem to exactly define the TO location. This occurrence is more evident when considering metal-poor
isochrones, whose morphology in the TO region is very much \lq{vertical}\rq.}, are in very good
agreement. In fact, at a given TO brightness, the  difference in the estimated age is of the order
of 1~Gyr or lower. It is also comforting that this difference among the various calibrations can be almost completely
explained by accounting for the different choices about the initial He content and the physical inputs.

\section{H-burning structures: the Red Giant Branch}

The RGB is one of the most prominent and well populated features in the CMD of   
stellar populations older than about $1.5 - 2$ Gyr.    
   
Since RGB stars are cool, reach high luminosities during their evolution, and their evolutionary 
timescales are relatively long, they provide a major contribution to the integrated   
bolometric magnitude and to integrated colors and spectra at wavelengths larger than about 900 nm of old   
distant, unresolved stellar populations (e.g. Renzini \& Fusi-Pecci~1988; Worthey~1994).   
A correct theoretical prediction of the RGB spectral properties and colors is thus of paramount importance 
for interpreting observations of distant stellar clusters and galaxies   
using population synthesis methods, but also for determining the ages of resolved stellar systems by 
means of isochron fitting techniques.   

Both the RGB location and slope in the CMD are strongly sensitive to the metallicity, and for this reason,
they are widely used as metallicity indicators.
  
The $I$-band brightness of the tip of the RGB (TRGB) provides a     
robust standard candle, largely independent of the stellar age and initial   
chemical composition, which can allow to obtain  
reliable distances out to about 10 Mpc using $HST$ observations   
(e.g., Lee, Freedman \& Madore~1993). Due to the lingering   
uncertainties on the empirical determination of the TRGB brightness   
zero point, RGB models provide an independent calibration of this important standard candle   
(Salaris \& Cassisi~1997, 1998). Theoretical predictions about the structural properties of RGB stars at the Tip
of the RGB play a fundamental role in determining the main evolutionary properties of their progeny: the
core He-burning stars during the Horizontal Branch (HB) evolutionary phase. In particular,     
HB luminosities (like the TRGB ones) are mostly determined by the value of the electron degenerate    
He-core mass ($M_{core}^{He}$) at the end of the RGB evolution.   
   
Predicted evolutionary timescales along the RGB phase play also a    
fundamental role in the determination of the initial He abundance 
of globular cluster stars through the R parameter    
(number ratio between HB stars and RGB stars   
brighter than the HB at the RR Lyrae instability strip level; see, e.g.   
Iben~1968a,  Salaris et al.~2004 and references therein), while   
an accurate modeling of the mixing mechanisms efficient  
in the RGB stars is necessary to correctly  
interpret spectroscopic observations of their surface chemical   
abundance patterns.    
  
The possibility to apply RGB stellar models to fundamental astrophysical problems crucially 
rely on the ability of theory to predict correctly:   
   
\noindent   
-- the CMD location (in $T_{\rm eff}$ and color) and extension    
(in brightness) of the RGB as a function of the initial chemical composition   
and age;   
   
\noindent   
-- the evolutionary timescales (hence the relative numbers of   
stars at different luminosities) all along the RGB;   
   
\noindent   
-- the physical and chemical structure of RGB stars, as well as   
their evolution with time.   
   
A detailed analysis of the existing uncertainties in theoretical RGB models, and of the level of
confidence in their predictions has been performed by Salaris, Cassisi \& Weiss (2002). In the following,
we will briefly review the main observational properties of the RGB such as its location and slope, the
bump of the luminosity function (LF) and the brightness of the Tip; discussing in some detail the main
sources of uncertainty in the corresponding theoretical predictions as well as the level of agreement
currently existing between theory and observations.

\subsection{The location and the slope of the RGB}

The main parameters affecting the RGB location and slope are: the EOS, the low-temperature opacity, the
efficiency of superadiabatic convection, the choice about the outer boundary conditions and the chemical
abundances.

\subsubsection{The EOS:}

Until a couple of years ago, the best available EOS was probably the OPAL one  (see previous discussion).  
However, its range of validity does not cover the electron degenerate cores   
of RGB stars and their cooler, most external layers, below 5000 K.   
RGB models computed with the OPAL EOS must employ some   
other EOS to cover the most external and internal stellar regions.
As a consequence, it was a common procedure to \lq{mix}\rq\ together EOSs provided by different authors in
order to have EOS tables suitable for the whole range of thermal conditions encountered by low-mass stars
from the H-burning stage to the more advanced evolutionary phase. However, there are some notable exception
about this as in the case of the stellar models computed by Vandenberg et al. (2000), Cassisi et al. (2003)
and Pietrinferni et al. (2004). In particular, in the case of the models presented by Cassisi et al. (2003)
and Pietrinferni et al. (2004), we take advantage by the use of the updated EOS computed by A. Irwin 
which consistently allows the computation of stellar models in both the H- and He-burning phases.

{\small
\begin{figure}
\hspace*{1.5cm}\psfig{figure=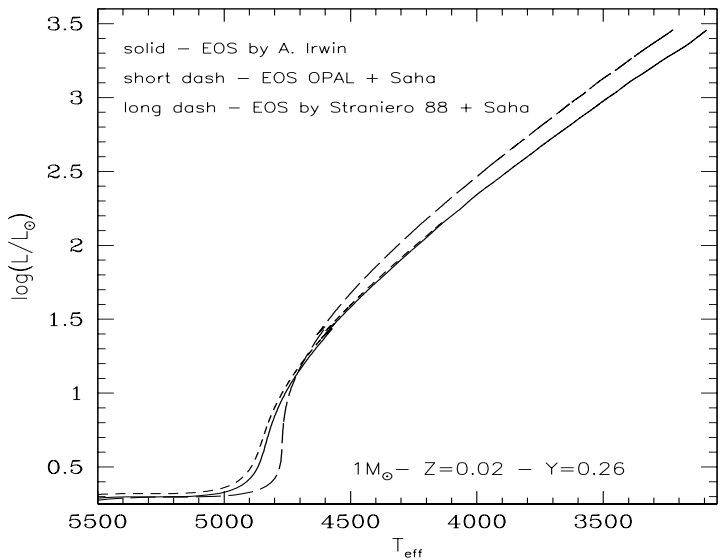,height=8truecm,width=10truecm}
\vspace*{-0.5cm}\caption{The RGBs of a $1M_\odot$ model computed by adopting different EOSs (see labels).}
\label{fig5}
\end{figure}
}
   
Till now, no detailed study exists highlighting the effect of the various EOS choices on the evolution \
and properties of RGB stars. In fig.~\ref{fig5}, the RGB of a $1M_\odot$ stellar structure computed
adopting different assumptions about the EOS is shown. One can notice that the models based on the OPAL EOS
and the EOS by A. Irwin are in very good agreement (this comparison is meaningful only for $T_{eff}$
larger than about 4500K for the reason discussed before); there is a significant change in the RGB slope
with respect the model based on the Straniero (1988) EOS supplemented at the lower temperature by a Saha
EOS. On average, there is a difference of about 100K between RGB models based of the two different EOSs.

\subsubsection{The low-temperature opacity:}

as shown by Salaris et al~(1993), it is the low-$T$ opacities which mainly determine the   
$T_{\rm eff}$ location of theoretical RGB models, while the    
high-$T$ ones - in particular those for temperature around $10^6K$ - enter in the determination of the    
mass extension of the convective envelope.
   
Current generations of stellar models employ mainly the low-$T$ opacity calculations 
by Alexander \& Ferguson~(1994) --  and in some cases the Kurucz~(1992) ones -- which are the most    
up-to-date computations suitable for stellar modelling, spanning a large range of initial chemical compositions.     
The main difference between these two sets   
of data is the treatment of molecular absorption, most notably the   
fact that Alexander \& Ferguson~(1994) include the effect of the $\rm H_{2}O$   
molecule. This last set of opacity accounts also for the presence of grains. These low-$T$ radiative 
opacity tabulations represent a remarkable improvement with respect the
old evaluations provided by Cox \& Stewart (1970) as far as it concerns the treatment of
molecules and grains. Although significant improvements are still possible as a consequence of a better
treatment of the different molecular opacity sources, we do not expect dramatic changes in the temperature
regime where the contribution of atoms and molecules dominate. Huge variation can be foreseen in the regime
($T<2000K$) where grains dominates the interaction between radiation and matter. These considerations
appear fully supported by the recent reanalysis of the low-$T$ opacities performed by Ferguson et al.
(2005).

{\small
\begin{figure}
\vspace*{-3cm}  
\hspace*{1.3cm}\psfig{figure=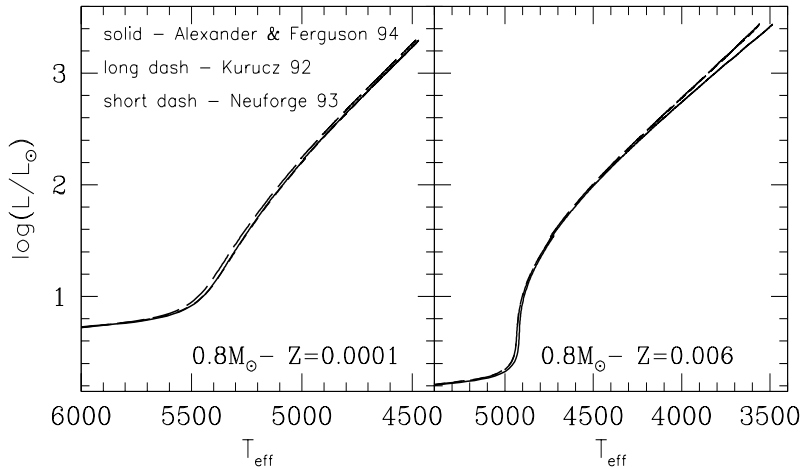,height=12truecm,width=11truecm}
\vspace*{-2.7cm}\caption{{\sl Left panel}: The RGBs of a $0.8M_\odot$ model computed by using different
prescriptions about the low-temperature radiative opacities and a metallicity $Z=0.0001$; {\sl right panel}: as the
left panel but for a metallicity $Z=0.006$.}
\label{fig6}
\end{figure}
}

Salaris \& Cassisi~(1996) have compared, at different   
initial metallicities, stellar models   
produced with these two sets of opacities (as well as with the    
less used Neuforge~1993 ones, which provide results almost   
undistinguishable from models computed with Kurucz~1992    
data), showing that a very good agreement exists   
when $T_{\rm eff}$ is larger than $\sim$4000 K as shown in fig.~\ref{fig6}.    
As soon as the RGB $T_{\rm eff}$ goes below this limit    
(when the models approach the TRGB   
and/or their initial metallicity is increased),    
Alexander \& Ferguson~(1994) opacities produce progressively    
cooler models (differences reaching values of the order of 100 K or more),    
due to the effect of the $\rm H_{2}O$   
molecule which contributes substantially to the opacity in this   
temperature range (see the right panel in fig.~\ref{fig6}). 

\subsubsection{The outer boundary conditions:}

the procedure commonly used in the current generation of stellar models is the integration of the
atmosphere by using a functional (semi-empirical or theoretical) relation between the temperature and the
optical depth ($T(\tau)$). Recent studies of the effect of using boundary   
conditions from model atmospheres are in V00 and   
Montalban et al.~(2001). In fig.~\ref{fig7} it is shown the effects on RGB stellar models 
of different $T(\tau)$ relations, namely, the Krishna-Swamy~(1966) solar T$(\tau)$  
relationship, and the gray one. One notices that RGBs computed with a gray  T$(\tau)$ are 
systematically hotter by $\sim$100 K. In the same Fig.~\ref{fig7}, 
we show also a RGB computed using boundary conditions from the Kurucz~(1992) model   
atmospheres, taken at $\tau$=10.  The three displayed RGBs, for consistency,   
have been computed by employing the same low-T opacities, namely the   
ones provided by Kurucz~(1992), in order to be homogeneous with the model atmospheres.   
The model atmosphere RGB shows a slightly different slope, crossing over the   
evolutionary track of the models computed with the 
Krishna-Swamy~(1966) solar T$(\tau)$, but the difference with respect to the latter stays always within   
$\sim \pm$50 K. 

{\small
\begin{figure}
\hspace*{1.5cm}\psfig{figure=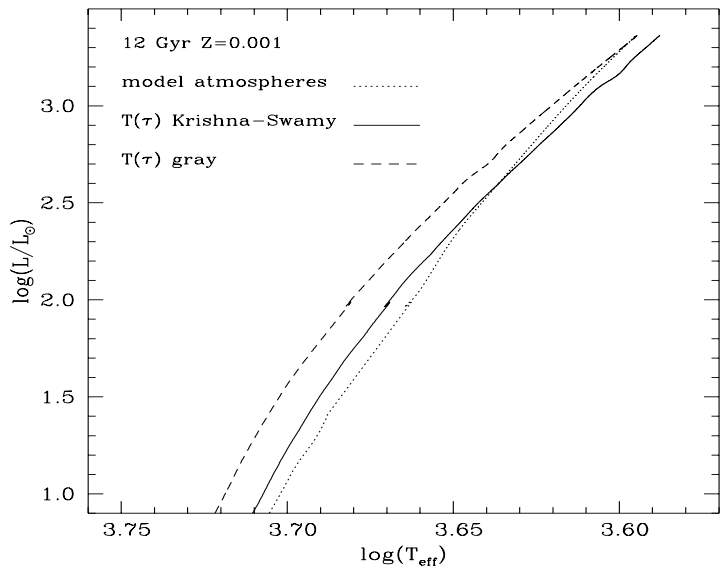,height=8truecm,width=10truecm}
\vspace*{-0.5cm}\caption{The RGB loci of a 11~Gyr isochron computed by adopting different
prescriptions about the outer boundary conditions (see labels) and a solar-calibrated mixing length.}
\label{fig7}
\end{figure}
}

Even if it is, in principle, more rigorous the use of boundary conditions provided by model atmospheres,
one has also to bear in mind that the convection treatment in the   
adopted model atmospheres (Montalban et al.~2001) is usually not the same as in the underlying   
stellar models (i.e., a different mixing length formalism and a
different value for the scale height of the convective motion is used).
 
\subsubsection{The chemical composition:}

as far as it concerns the helium abundance, the evolutionary properties of RGB stars, at least concerning
their effective temperatures, are not strongly affected by different assumptions on the initial He content.
This occurrence is due to the combination of two different reasons: 1) stellar matter opacity does not
strongly depend on the He abundance, 2) the initial He abundance for old, stellar systems such as GGCs is well
constrained and variations larger that $\sim0.02-0.03$ are unrealistic.
On the contrary,  the abundance of heavy elements is one of the parameters which most
affects the RGB morphology: any increase of $Z$ produces a larger
envelope opacity and, in turn, a more extended envelope convection
zone and a cooler RGB. The strong dependence of the RGB effective
temperature on the metallicity makes the RGB one of the most
important metallicity indicators for stellar systems. An important issue is the
dependence of the shape and location  of the RGB on the
distribution of the metals: different heavy elements have
different ionization potentials, and provide different contribution to the envelope opacity. The abundance of
low ionization potential elements such as Mg, Si, S, Ca, Ti and Fe
strongly influences the RGB effective temperature, through their
direct contribution to the opacity due to the formation of
molecules such as TiO which strongly affects the stellar spectra
at effective temperatures lower than $5000-6000K$, and through the
electrons released when ionization occurs, which affect the
envelope opacity via the formation of the H$^-$ ion $-$ one of the
most important opacity sources in RGB structures. As an example, a 
change of the heavy elements mixture from a scaled solar 
one to an $\alpha$-element enhanced
distribution with the same iron content, produces a larger envelope
opacity and the RGB becomes cooler and less steep: the change
in the slope being due to the increasing contribution of molecules
to the envelope opacity when the stellar effective temperature
decreases along the RGB.

\subsubsection{The treatment of superadiabatic convection:}

in section~\ref{msto} we already noticed that the value of $\alpha_{\rm MLT}$   
is usually calibrated by reproducing the solar $T_{\rm eff}$, and this   
solar-calibrated value is then used for stellar models of different masses and along different evolutionary phases,
including the RGB one. 
The adopted procedure guarantees that the models always predict correctly the $T_{\rm eff}$  
of at least solar type stars. However, 
the RGB location is much more sensitive to the value of $\alpha_{\rm MLT}$   
than the main sequence. This is due to the evidence that along the RGB the extension (in radius) of the 
superadiabatic layers - as a consequence of the much more
expanded configuration achieved by the star - is quite larger when compared with the MS evolutionary phase.
Therefore, it is important to verify that a solar $\alpha_{\rm MLT}$ is always suitable also for RGB stars of   
various metallicities.   
   
An independent way of calibrating $\alpha_{\rm MLT}$ for RGB stars is to compare empirically determined 
RGB $T_{\rm eff}$ values for galactic GCs with theoretical models of the appropriate chemical   
composition (see also Salaris \& Cassisi~1996, Vandenberg, Stetson \& Bolte~1996 and references therein).   
In fig.~\ref{fig8}, as an example taken from the literature, we show a comparison between  the  
$T_{\rm eff}$ from Frogel, Persson \& Cohen~(1983) for a sample of GCs and the $\alpha$-enhanced models by SW98. 
For a detailed discussion of how current empirical uncertainties on the GGCs distance scale, metallicity scale
and RGB temperature scale affects the comparison shown in fig.~\ref{fig8}, we refer to Salaris et al. (2002).
The results shown in this figure (recently confirmed also by Vandenberg et al.~2000, by using their own updated
set of RGB stellar models) seem to suggest that the solar $\alpha_{\rm MLT}$ value 
is {\sl a priori} adequate also for RGB stars (but see also the discussion in Salaris et al.~2002).

{\small
\begin{figure}
\vspace*{-1cm}  
\hspace*{1.5cm}\psfig{figure=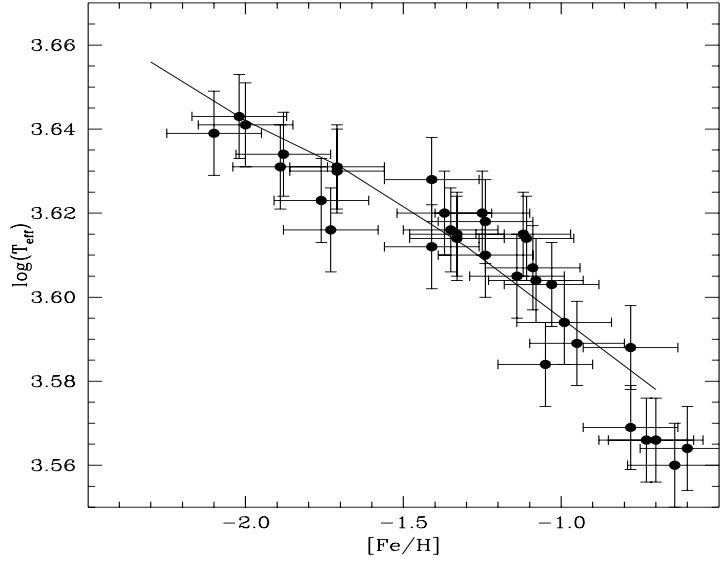,height=8truecm,width=10truecm}
\vspace*{-0.5cm}\caption{Empirical estimates of the RGB effective temperature in a sample of GGCs
as a function of [Fe/H]. The solid line represents the theoretical calibration provided by Salaris \& Weiss (1998) (see text for
more details). The stellar models have been computed by adopting a solar-calibrated mixing length.}
\label{fig8}
\end{figure}
}
  
This notwithstanding, a source of concern about an {\sl a priori} assumption of a solar $\alpha_{\rm MLT}$   
for RGB computations comes from the fact that recent models from various authors, all using a suitably     
calibrated solar value of $\alpha_{\rm MLT}$, do not show the same RGB temperatures.    
This means that -- for a fixed RGB temperature scale --   
the calibration of $\alpha_{\rm MLT}$ on the empirical $T_{\rm eff}$ values   
would not provide always the solar value.   
Figure~\ref{fig9} displays several isochrones produced by different  
groups (see labels and figure caption), all computed with the same initial chemical composition, same opacities,    
and the appropriate solar calibrated values of $\alpha_{\rm MLT}$: the Vandenberg et al. (2000) and Salaris
\& Weiss (1998) models are identical, the Padua ones (Girardi et al. 2000) are systematically hotter 
by $\sim$200 K, while the $Y^2$ ones (Yi et al.~2001) have a different shape.  
This comparison shows clearly that if one set of MLT solar calibrated RGBs  
can reproduce a set of empirical RGB temperatures, the others cannot,  
and therefore in some case a solar calibrated $\alpha_{\rm MLT}$  
value may not be adequate.  
The reason for these discrepancies must be due to some    
difference in the input physics, like the EOS and/or the boundary conditions,    
which is not compensated by the solar recalibration of $\alpha_{\rm MLT}$.   

{\small
\begin{figure}
\hspace*{1.5cm}\psfig{figure=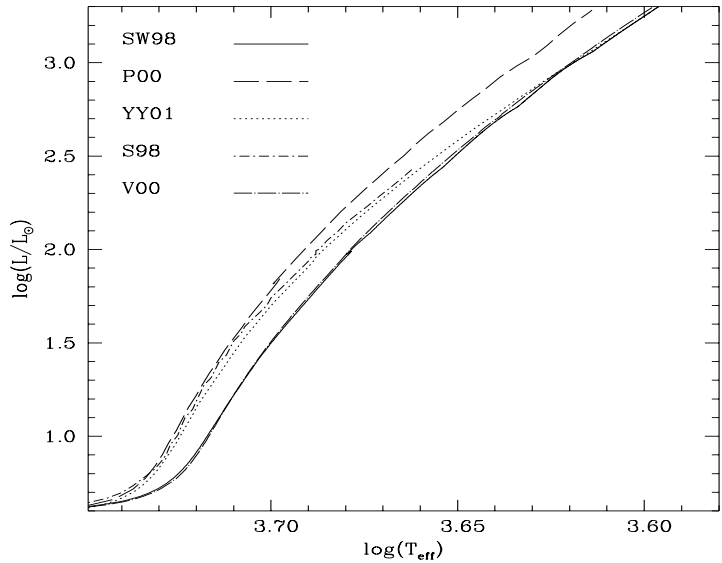,height=8truecm,width=10truecm}
\vspace*{-0.5cm}\caption{RGB isochrones computed with a solar-calibrated MLT, for a scaled solar metal 
mixture and  an age of 12~Gyr, as provided from different authors: Girardi et al. (2000, P00), 
Yonsei-Yale models (Yi et al.~2001, YY01), VandenBerg et al. (2000, V00), Salaris \& Weiss (1998, SW98),
and the FST models by Silvestri et al. (1998, S98).
}
\label{fig9}
\end{figure}
}

To illustrate this point   
in more detail, we show in fig.~\ref{fig10} two evolutionary tracks for a $1M_\odot$ stellar model
with solar chemical composition. The only difference between them is the treatment of the boundary   
conditions. Two different T($\tau$) relationships,    
namely gray and Krishna-Swamy~(1966), have been adopted.   
The value of $\alpha_{\rm MLT}$ for the two models has been    
calibrated in each case, in order to reproduce the Sun,    
and in fact the two tracks    
completely overlap along the main sequence, but the RGBs show a
difference of the order of 100 K.
  
This occurrence clearly points out the fact that one  
cannot expect the same RGB $T_{\rm eff}$ from solar calibrated models  
not employing exactly the same input physics. The obvious conclusion  
is that it is always necessary to compare RGB models with observations  
to ensure the proper calibration of $\alpha_{\rm MLT}$ for RGB stars.  

{\small
\begin{figure}
\vspace*{0.6cm}  
\hspace*{1.5cm}\psfig{figure=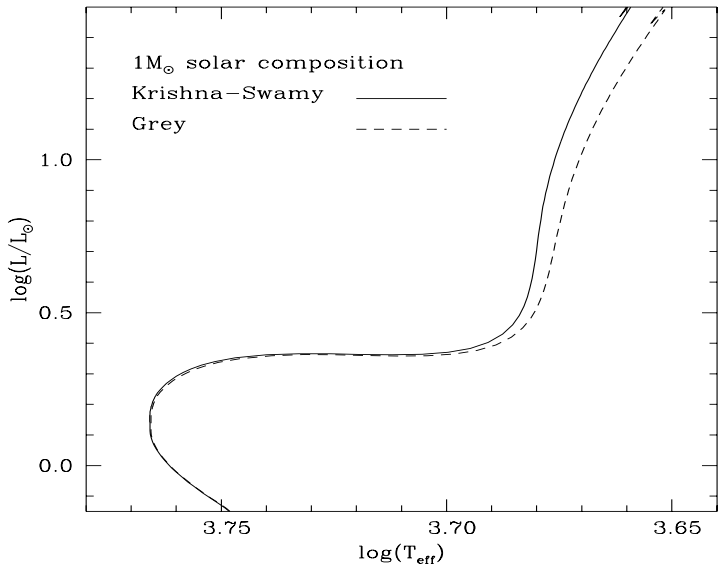,height=8truecm,width=10truecm}
\vspace*{-0.5cm}\caption{The evolutionary tracks in the H-R diagram of two models of $1M_\odot$,
computed by adopting a solar-calibrated MLT, but using two different $T(\tau)$ relations for fixing the
outer boundary conditions.}
\label{fig10}
\end{figure}
}
  
\subsection{The bump of the RGB luminosity function}
\label{bump}

The RGB luminosity function (LF), i.e. the number of stars per brightness bin among the RGB as a function
of the brightness itself, of GGCs is an important tool to test the chemical stratification inside
the stellar envelopes (Renzini \& Fusi Pecci 1988).
The most interesting feature of the RGB LF is the occurrence of
a local maximum in the luminosity distribution of RGB stars, which
appears as a bump in the differential LF, and as a change in the slope
of the cumulative LF. This feature is caused by
the sudden increase of H-abundance left over
by the surface convection upon reaching its maximum inward extension at
the base of the RGB (\emph{first dredge up}) (see Thomas~1967 and Iben~1968b). 
When the advancing H-burning shell encounters this
discontinuity, its efficiency is affected (sudden increase of the
available fuel), causing a temporary drop of the surface luminosity. After some time the
thermal  equilibrium is restored and the surface luminosity starts
to increase again. As a consequence, the stars cross the same
luminosity interval three times, and this occurrence shows up as a characteristic
peak in the differential LF of RGB stars. Moreover, since the H-profile before and after the discontinuity
is different, the rate of advance of the H-burning shell changes when the discontinuity is crossed, thus causing 
a change in the slope of the cumulative LF.

The brightness of the RGB bump is therefore related to the location of
this H-abundance discontinuity, in the sense that the deeper the chemical
discontinuity is located, the fainter is the bump luminosity. As a consequence, any physical inputs
and/or numerical assumption adopted in the computations, which affects the maximum
extension of the convective envelope at the \emph{first dredge up}, strongly affects the bump
brightness.
A detailed analysis of the impact of different physical inputs on the predicted RGB bump luminosity
can be found in Cassisi \& Salaris (1997) and Cassisi, Salaris \& Degl'Innocenti (1997) and it will not
be repeated. However, it is worth noting that a comparison between the predicted bump luminosity and 
the observations allows a direct check of how well theoretical models for RGB stars
predict the extension of convective regions in the stellar envelope and, then provide a
plain evidence of the reliability of current evolutionary framework (Valenti, Ferraro \& Origlia 2004).

{\small
\begin{figure}
\vspace*{-0.8cm}  
\hspace*{1.5cm}\psfig{figure=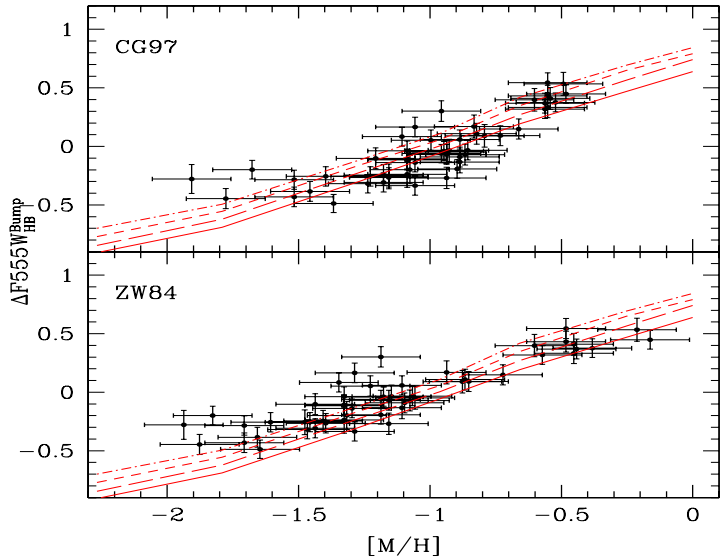,height=7.8truecm,width=10truecm}
\vspace*{-0.4cm}\caption{Comparison between empirical measurements of the brightness difference between
the RGB bump and the ZAHB (in the HST $F555W$ filter) and theoretical prescriptions as given by Pietrinferni et al.~(2004)
as a function of the global metallicity by adopting both the Zinn \& West (1984, ZW84) and Carretta \& Gratton (1997, CG97) 
metallicity scales (see Riello et al.~2003 for more details). The theoretical predictions are plotted for four different
cluster ages, namely, from bottom to top, 10~Gyr, 12~Gyr, 14~Gyr and 16~Gyr.}
\label{fig11}
\end{figure}
}

In this context it is worth noting that, for each fixed global metallicity, the theoretical predictions about the 
Bump luminosity provided by Bergbush \& Vandenberg (2001) are in fine agreement - within
$\approx0.05$ mag -, with the values given by Cassisi \& Salaris (1997). This is a plain evidence of the
fact that current, updated canonical stellar models do agree to a significant level about this relevant 
evolutionary feature.

However, when comparing theory with observations, one needs a preliminary estimate of both the cluster metallicity and
distance. Current uncertainty in the GGC metallicity scale strongly reduces our capability to 
constrain the plausibility of the theoretical framework 
(see the discussion in Bergbusch \& Vandenberg 2001), and for such reason,
it has became a common procedure to use simultaneously all available metallicity scales (see Riello et al. 2003). 
Another critical issue is related to the need of knowing the cluster distance, whose accuracy could strongly hamper the
possibility of a meaningful comparison between theory and observations. Following the early prescription 
provided by Fusi Pecci et al.~(1990), the observed magnitude
difference between the RGB bump and the HB at the RR Lyrae instability
strip (\dvbump) is usually adopted in order to test the theoretical
predictions for the bump brightness. This quantity presents several advantages from the
observational point of view (see Fusi Pecci et al.~1990, and
Salaris et al.~2002) and it is empirically well-defined
because it does not depend on a previous knowledge of the cluster distance and reddening. However, on the theoretical
side, one should keep in mind that such comparison requires the use of a theoretical prediction about the Horizontal
Branch brightness which is a parameter still affected by a significant uncertainty (see section~\ref{zahb}). Nevertheless, 
empirical estimates about the \dvbump parameter have been extensively compared with theoretical predictions (Riello et al. 2003 and 
references therein). Figure~\ref{fig11} shows the results of the comparison performed by Riello et al. (2003): 
even though a qualitative agreement between theory and observations of \dvbump\ does exist, a more definitive 
assessment of the confidence level appears clearly hampered by the not negligible uncertainties still affecting
both the cluster [Fe/H] and [$\alpha$/Fe] estimates. In conclusion, it is realistic to consider that due to lingering
uncertainties on the (theoretically determined!) HB brightness and the GGCs metallicity scale, there is the
possibility of a discrepancy between theory and observation about the \dvbump parameter at the level of $\sim0.20$ mag.

Before concluding this section, we wish to notice that the RGB LF bump provides other important
constraints besides brightness for checking the accuracy of theoretical RGB models. 
More in detail, both the shape and the location of the bump along the RGB LF can be used for investigating on
how \lq{steep}\rq\ is the H-discontinuity left over by envelope convection at the \emph{first dredge
up}. So these features appear, potentially, a useful tool for investigating on the efficiency of non-canonical
mixing at the border of the convective envelope (Cassisi, Salaris \& Bono 2002) able to partially smooth the chemical
discontinuity. In addition, since the evolutionary rate along the RGB is strongly affected by any change in the
chemical profile, it is clear that the star counts in the bump region can provide reliable information about the size
of the jump in the H profile left over by envelope convection after the \emph{first dredge up}. This issue,as well as the level
of agreement between theory and observations, has been investigated by Bono et al. (2001) and Riello et al. (2003).

\subsection{Star counts along the RGB}

The number of stars in any given bin of the RGB LF is determined by the local evolutionary rate so the comparison
between empirical and theoretical RGB LF represents a key test for the accuracy of the predicted RGB timescales (see
Renzini \& Fusi Pecci 1988). In addition, there are many more reasons for which to investigate the RGB star counts is
quite important, for instance: i) being the RGB stars among the brightest objects in a galaxy, their number has a
strong influence on the integrated properties of the galactic stellar population; ii) the number ratio between RGB and
stars along the Asymptotic Giant Branch (AGB) can be used to constrain the Star Formation History of a galaxy (Greggio
2000).

{\small
\begin{figure}
\hspace*{1.5cm}\psfig{figure=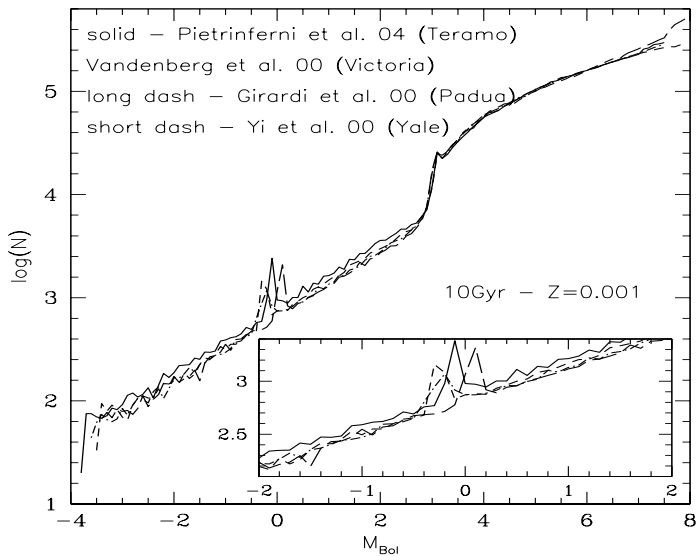,height=8truecm,width=10truecm}
\vspace*{-0.5cm}\caption{Comparison among various luminosity functions as given by different authors (see labels). The number of MS
stars is the same in all cases. The inset shows in more detail the location of the different LFs during the RGB evolutionary stage.}
\label{fig12}
\end{figure}
}

A recent, investigation of the accuracy of theoretical RGB LFs has been performed by Zoccali \& Piotto (2000)
by adopting a large database of GGC RGB LFs. The main outcome of their analysis was the evidence of, on average,
a good agreement, on the whole explored
metallicity range, between observations and the theoretical predictions, available at that time. However, more recently, Gallart, Zoccali \& Aparicio (2005) have 
reanalyzed this issue and have
noticed  that for a fixed number of MS stars, the number of RGB stars as predicted by different sets of evolutionary
models are not in good agreement: in the smooth part of the RGB LF the different models show differences as large as
0.15 dex, that correspond to a factor of $\sim1.4$ of difference in the number counts. This result is shown in
fig.~\ref{fig12}, where LFs from different authors and for various metallicities are plotted.

\subsection{The Tip of the RGB}

As a consequence of the H-burning occurring in the shell, the mass size of the He core ($M_{core}^{He}$) grows. 
When $M_{core}^{He}$ reaches about 0.50 $M_{\odot}$ (the precise   
value depends weakly on the total mass of the star for structures less massive than about $1.2M_\odot$, i.e. older than about
4-5Gyr, being more sensitive    
to the initial chemical composition), He-ignition occurs in the   
electron degenerate core. This process is the so called He-flash, that terminates the RGB phase   
by removing the electron degeneracy in the core and, driving the star onto its Zero Age Horizontal Branch (ZAHB) location, that   
marks the start of quiescent central He-burning plus shell H-burning. The brightest point along the RGB, that marks the
He ignition through the He flash is the so-called Tip of the RGB (TRGB).

The observational and evolutionary properties of RGB stars at the TRGB play a pivotal role in current stellar astrophysical
research. The reasons are manifold: i) the mass size of the He core at the He flash fixes not only the TRGB brightness but
also the luminosity of the Horizontal Branch, ii) the TRGB brightness is one of the most important primary distance indicators.

More in detail, the reasons which make the TRGB brightness a quite suitable standard candle are the following: the TRGB
luminosity that is a strong function of the He core mass at the He flash, is weakly dependent on the stellar mass, and therefore on the 
cluster age over a wide age interval. This is due to the already mentioned evidence that the value of
$M_{core}^{He}$ at the He-flash is fairly  
constant over large part of the low-mass star range. However,
$M_{core}^{He}$ decreases for increasing   
metallicity, while the TRGB brightness increases due to the increased   
efficiency of the H-shell, which compensates for the reduced core mass\footnote{We recall that 
the brightness of the subsequent ZAHB phase   
follows the behaviour of $M_{core}^{He}$, decreasing for increasing    
metallicity.}.   

{\small
\begin{figure}
\vspace*{-0.5cm}  
\hspace*{1.5cm}\psfig{figure=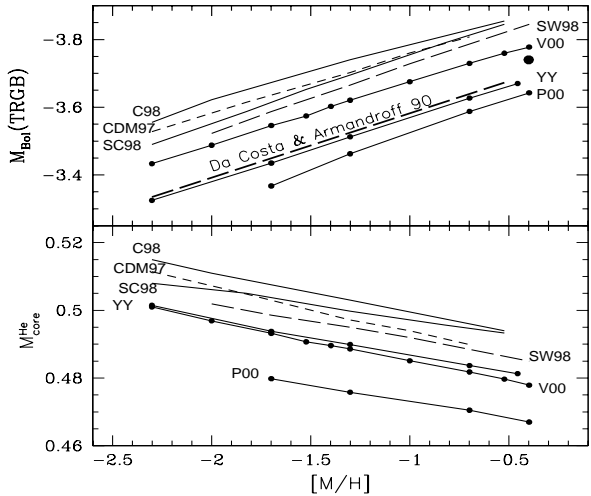,height=8.0truecm,width=9truecm}
\vspace*{-1.0cm}\caption{{\sl Upper panel}: the trend of the bolometric magnitude of the TRGB
as a function of the metallicity as provided by different authors: Cassisi et al.~(1998, C98), Caloi, D'Antona \& Mazzitelli~(1997,
CDM97), Salaris \& Cassisi~(1998, SC98), Salaris \& Weiss~(1998, SW98), Vandenberg et al.~(2000, V00), Yi et al.~(2001, YY), Girardi
et al.~(2000, P00). The full circle marks the location of the prescription provided by Salasnich et al.~(2000). The semi-empirical
calibration provided by Da Costa \& Armandroff (1990) is also shown. {\sl Lower panel}: as the upper panel but for the mass size of
the core at the RGB.}
\label{fig13}
\end{figure}
}

Luckily enough, the value of $\rm M_{I}^{TRGB}$ - the I-cousin band magnitude of the TRGB - appears to be very weakly sensitive to the   
heavy element abundance  (Lee, Freedman \& Madore~1993 and Salaris \& Cassisi~1997): for $[M/H]$ ranging between    
$-$2.0 and $-$0.6, $\rm M_{I}^{TRGB}$ changes by less than 0.1 mag. This lucky occurrence stands from the evidence that    
$\rm M_{bol}^{TRGB}$ is proportional to $\sim -$0.18[M/H], while $\rm BC_{I}$ is   
proportional to $\sim -0.14[M/H]$. Therefore, the slope of the $\rm BC_{I}-[M/H]$ relationship is quite similar to the slope of    
the $\rm M_{bol}^{TRGB}-[M/H]$ relationship, and since    
$\rm M_{I}^{TRGB}$=$\rm M_{bol}^{TRGB}- \rm BC_{I}$, it results that $\rm M_{I}^{TRGB}$   
is almost independent of the stellar metal content.   

As far as it concerns the uncertainties affecting theoretical predictions about the TRGB brightness, it is clear that, being the
TRGB brightness fixed by the He core mass, any uncertainty affecting the predictions about
the value of $M_{core}^{He}$ immediately translates in an error on $\rm M_{bol}^{TRGB}$.
An exhaustive analysis of the physical parameters which affects the estimate of $M_{core}^{He}$ provided by stellar models has been
provided by Salaris et al. (2002, but see also Castellani \& Degl'Innocenti 1999): the main result of these analyses was that the
physical inputs which have the largest impact in the estimate of $M_{core}^{He}$ are the efficiency of atomic diffusion and the
conductive opacity.

\subsubsection{Atomic diffusion:} Castellani \& Degl'Innocenti (1999) have clearly shown that to change by a factor of 2 the
efficiency of microscopic diffusion (a realistic estimate of current uncertainty affecting the efficiency of this
process) causes a change of about $(-0.002/+0.004)M_\odot$ in the value of $M_{core}^{He}$ (the He core mass increasing when the atomic
diffusion efficiency is increased).

\subsubsection{Conductive opacities:}
since the conductive transport efficiency regulates the thermal state of the electron degenerate He core,    
a reliable estimate of the conductive opacities is fundamental for deriving the   
correct value of the He-core mass at the He-flash. As a general rule, higher conductive opacities cause a    
less efficient cooling of the He-core and an earlier He-ignition (i.e., at a lower core mass).   
   
Until few years ago, only two choices were available, neither of   
which is totally satisfactory: the analytical relation provided by Itoh et al.~(1983, I83),    
or the old Hubbard \& Lampe (1969, HL) tabulation.   
As pointed out by Catelan, de Freitas Pacheco \& Horvath~(1996),   
the most recent results by I83 are an improvement over the older HL  
ones, but their range of validity does not cover the He-cores of RGB stars. When using the I83 conductive opacity, 
Castellani \& Degl'Innocenti~(1999) found an increase - with respect to the models based on the HL opacity - by 0.005$M_{\odot}$ of   
$M_{core}^{He}$ core at the He-flash for a 0.8$M_{\odot}$ model with initial   
metallicity Z=0.0002, while   
in case of a 1.5$M_{\odot}$ star with solar chemical composition the   
increase amounts to 0.008$M_{\odot}$ (Castellani et al.~2000). 
Quite recently, new estimates for the conductive opacity has been provided by Potekhin (1999). This new set represents
a significant improvement (both in the accuracy and in the range of validity) with respect to previous estimates. It is worth noting
that RGB stellar models based on these new conductive opacities provided He core masses at the He-ignition
whose values are intermediate between those provided by the previous conductive opacity estimates (although more similar to the determinations based
on the Itoh et al.~(1983) ones). However, as noticed by Potekhin (1999) and further emphasized by Catelan (2005), not even these
new conductive opacity fully covers the thermal conditions characteristic of electron degenerate cores in low-mass, metal-poor
stars. So it is evident that additional work in this direction is strongly encouraged.

As far as it concerns the EOS, a preliminary investigation on the effect
of different EOS choices has been performed by Vandenberg \& Irwin~(1997) and more recently by Cassisi et al. (2003): it has been
noticed that, when the adopted EOS accounts for all the different physical processes at work in the dense core of RGB stars, the
residual uncertainty on the value of $M_{core}^{He}$ can be small. The impact
of current uncertainties on the relevant nuclear reaction rates as the one corresponding to the $3\alpha$ process has been
recently investigated by Weiss et al. (2005, but see also Brocato et al. 1998) with the result that present uncertainty on the relevant rate has no significant
influence on theoretical predictions about the TRGB.

When, one considers as \lq{standard}\rq\  a model accounting for standard atomic diffusion, current uncertainties in diffusion
efficiency and conductive opacity can globally contribute to an uncertainty on the He core mass of the order of
$\sim0.01M_\odot$. It can be useful to briefly remember that, since $\partial\log(L_{TRGB})/\partial{M_{core}^{He}}\approx4.7$, this
uncertainty immediately translates in an error of about $\sim0.10$mag in the bolometric magnitude of the TRGB. In our
belief, this is a realistic estimate of current uncertainty affecting theoretical predictions on this relevant feature.

We show in fig.~\ref{fig13} the comparison of the most recent results concerning     
the TRGB bolometric magnitude and $M_{core}^{He}$ at the He-flash; the   
displayed quantities refer to a 0.8$M_{\odot}$ model and various   
initial metallicities (scaled solar metal distribution).   
  
There exists fair agreement among the various predictions   
of the $\rm M_{bol}^{\rm TRGB}$ metallicity dependence, and all 
the $\rm M_{bol}^{\rm TRGB}$ values    
at a given metallicity are in agreement within $\sim 0.10$ mag, with the exception    
of the Padua models (Girardi et al.~2000) and the Yale ones (Yi et al.~2001), which appear to be underluminous with   
respect to the others.    
As for the Padua models this difference follows from their smaller $M_{core}^{He}$ values;   
it is worth noticing that the recent models by Salasnich et   
al.~(2000), which are an update of the Padua ones, provide    
brighter $\rm M_{bol}^{\rm TRGB}$, similar to the results by Vandenberg et al. (2000).
In case of the Yale models, the result is surprising since the fainter TRGB luminosity cannot be explained
by much smaller $M_{core}^{He}$ values, since this quantity is very similar to, for instance, the results given by Vandenberg et
al.~(2000).
When neglecting the Padua and Yale models, the 0.1 mag   
spread among the different TRGB brightness estimates can be interpreted in terms of differences in the adopted physical inputs
such as for instance the electron conduction opacities.

{\small
\begin{figure}
\vspace*{-0.5cm}  
\hspace*{1.5cm}\psfig{figure=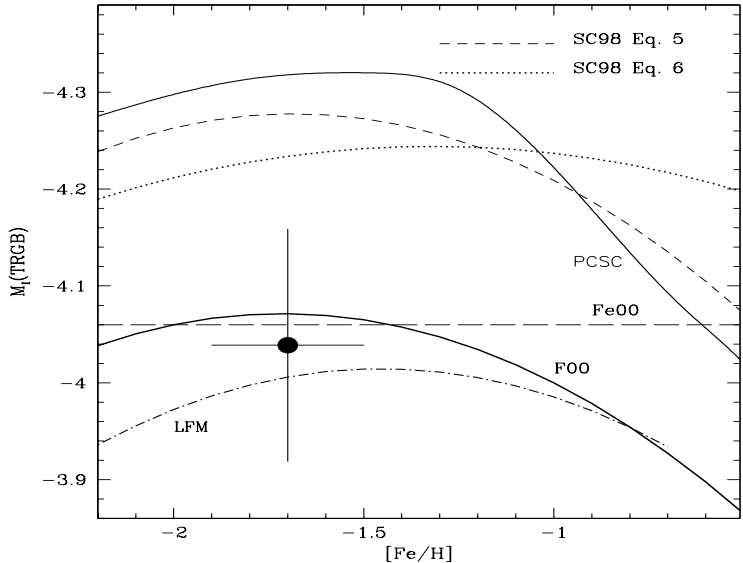,height=8truecm,width=10truecm}
\vspace*{-0.5cm}\caption{A comparison among theoretical calibrations of the I-cousin band brightness of the TRGB as provided by
Pietrinferni et al.~(2004, PCSC) and Salaris \& Cassisi~(1998, SC98, see their equations~5 and 6), and empirical or semi-empirical ones as
given by Lee et al.~(1993, LFM), Ferrarese et al.~(2000, Fe00) and Ferraro et al.~(2000, F00). The full circle with the error bars
corresponds to the empirical calibration provided by Bellazzini et al.~(2001).}
\label{fig14}
\end{figure}
}

Due to the large relevance of the TRGB as standard candle, it is worthwhile showing a comparison between (some) theoretical
predictions about the I-cousin band magnitude of the TRGB and empirical calibrations. This comparison is displayed in
fig.~\ref{fig14}, where we show also the recent empirical calibration provided by Bellazzini et al. (2001) based on the GGC
$\omega$~Cen. In this plot, we have shown different calibrations of $M_I^{TRGB}$ as a function of the metallicity based on our own
stellar models. These calibration are about $0.15-0.20$ mag brighter than the most recent, empirical ones. When considering also
various theoretical calibrations as those displayed in fig.~\ref{fig14}, one notices that these updated calibrations of $M_I^{TRGB}$ 
are within $\sim1.5\sigma$ of the calibration provided by Bellazzini et al. (2001).

In this context, it can be useful to remember that in order to derive this calibration a bolometric correction scale for the 
$I-$band has to be used (see Salaris \& Cassisi 1998) that, as it is well known, can be affected by large uncertainty. Therefore, it
appears quite difficult at this time to disentangle the contribution to the global discrepancy between theoretical and empirical
calibration, due to current uncertainty in the $BC_I$ scale from that associated to present uncertainties in stellar RGB
models. In order to illustrate better this point, we show in fig.~\ref{fig15} a comparison between our theoretical calibration
in different photometric planes and the corresponding empirical ones provided by Bellazzini et al. (2004): the evidence that the same
theoretical calibration does not work properly for the I-cousin band while providing a very good match in the near-infrared bands
strongly shed light on the importance of an accurate and critical analysis on the uncertainties affecting the $BC$ scales for the
various photometric bands.

{\small
\begin{figure}
\vspace*{-0.5cm}  
\hspace*{1.5cm}\psfig{figure=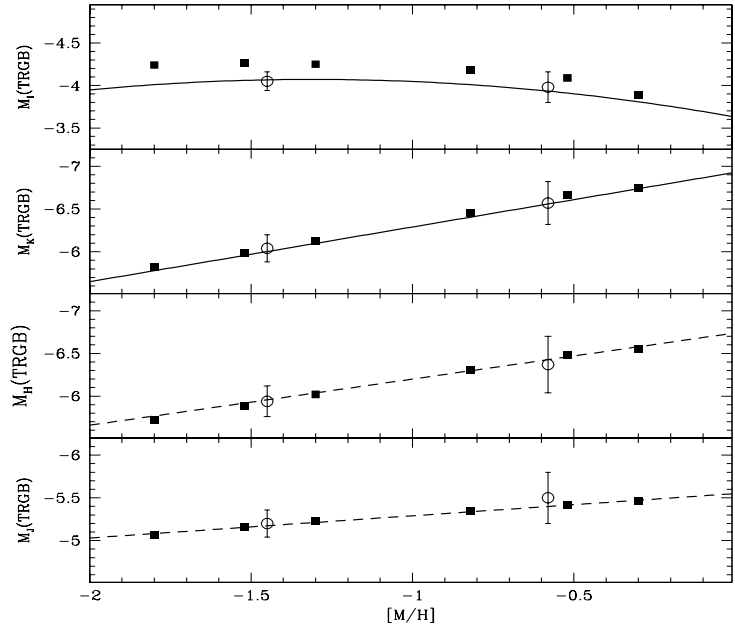,height=9truecm,width=10truecm}
\vspace*{-0.5cm}\caption{A comparison between empirical data (open circles) and theoretical prescriptions concerning the brightness of the TRGB
in different photometric bands. In the two lower panels the dashes lines show a linear regression to the theoretical data, while
in the two upper panel, the solid lines represent empirical calibrations (see Bellazzini et al.~2004 for more details).}
\label{fig15}
\end{figure}
}

\section{The He-burning structures}
\label{zahb}

The Horizontal Branch is one of the most important evolutionary sequences in the CMD. The reasons for this relevance are manifold:
1) the brightness of the RR Lyrae stars and more, in general, the brightness of the HB, is the traditional distance indicator 
for metal-poor stellar populations; 2) the number of stars observed along this branch (a quantity tightly related with the core 
He-burning lifetimes) enters in the R parameter definition (Buzzoni et al. 1983), the most important He indicator for old
stellar systems; 3) the HB morphology is related to the long-standing, and still unsettled, problem of the \lq{second parameter}\rq\ in the
galactic GC system.

From a theoretical point of view, although we know well for a long time the structural and evolutionary properties of HB stars,
we can not yet be fully confident in the theoretical predictions concerning this evolutionary phases, at least as far as it
concerns the luminosity and the evolutionary lifetime. This is simply due to the evidence that the evolutionary properties of HB
stars strongly depend on all the physical processes at work during the early RGB phase. Therefore, the uncertainties affecting
the physical scenario used for computing  H-burning structures appear, in some sense, amplified when considering HB stellar
models which are, in addition, affected by other sources of uncertainty as the rates of the He-burning processes and the
efficiency of mixing at the border of the convective core. 

This topic has been accurately analyzed by Castellani \& Degl'Innocenti (1999), which have investigated  the sensitivity of HB
luminosities to the uncertainties affecting the various physical inputs. The errors affecting the HB evolutionary lifetimes have
been extensively reviewed by Brocato et al. (1998) and Cassisi et al. (1998, 2003).

\subsection{The luminosity of the Horizontal Branch}

It is well known that the bolometric luminosity of a ZAHB structure is governed by two parameters: mostly the He core mass and,
to a minor extent, the chemical stratification of the envelope. On the basis of this consideration, one can easily realize
that the whole set of uncertainties affecting the value of $M_{core}^{He}$ at the TRGB directly affect also the HB brightness. 
When considering only as sources of error: the atomic diffusion efficiency and the different choices about
the conductive opacity; one derives that the visual absolute magnitude of the HB is uncertain at the level of $\approx0.10$ mag,
corresponding to a $\sim10$\% uncertainty on this relevant feature. A more detailed analysis accounting for all error sources
shows that the stellar evolution theory predicts the ZAHB luminosity with an uncertainty on $M_V$ of $-0.06/+0.11$
mag\footnote{This estimate does not account for any contribution to the error budget coming from the adopted $BC_V$ scale.}.

In fig.~\ref{fig16}, a comparison among different, updated, theoretical predictions is shown. For the sake of comparison, we
display in the same plot the semi-empirical ZAHB brightness estimates provided by De Santis \& Cassisi (1999).

\subsection{The core He-burning lifetime}
\label{hbtime}

As for any evolutionary phase, the lifetime of the core He-burning phase - for a given total mass - does depend on the \lq{speed}\rq\ at which the nuclear
processes occur, i.e., on the nuclear reaction rates, and on the amount of available fuel, i.e., in the case of burning occurring
in a convective core, on the location of the outer convective boundary. Really, the uncertainty on the HB evolutionary lifetimes
is dominated by the uncertainty on nuclear reaction rates and by, to a minor extent (see below) from the not-well known,
efficiency of convective processes.

Concerning the reaction rates,  it is worth emphasizing that the ${\rm {^{12}C}(\alpha,\gamma){^{16}O}}$ 
reaction is, together with the triple$-\alpha$ process, the most
important among those involved in the He-burning. This occurrence being due to the evidence that: i)
its nuclear cross-section strongly affects the C/O ratio in the core of carbon-oxygen white dwarfs and, 
in turn, their cooling times; ii) when the abundance of He inside the convective core is significantly reduced, 
the ${\rm {^{12}C}(\alpha,\gamma){^{16}O}}$ reaction becomes strongly competitive with the $3\alpha$ reactions 
(which need three $\alpha$ particles) in supplying the nuclear energy budget. This means that the cross-section
of this nuclear process has a huge impact on the core He-burning lifetime as well as on the chemical stratification in the core at
the central He exhaustion.

Unfortunately, this reaction has a resonance and a very low cross-section at low energies, and so the
nuclear parameters are difficult to measure experimentally or to calculate by theoretical analysis. As a consequence,
an uncertainty of a factor of 2 is reasonable for this nuclear reaction rate (Caughlan \& Fowler 1988, 
but see also the recent analysis by Kunz et al~2002).

The uncertainty on the ${\rm {^{12}C}(\alpha,\gamma){^{16}O}}$ reaction rate strongly affects the HB lifetime:
according to Brocato et al. (1998) the HB lifetime change correlates with a variation of the rate as
$\Delta{t_{HB}}/t_{HB}\sim0.10\Delta\sigma_{^{12}C}/\sigma_{^{12}C}$ (see also Zoccali et al.~1999).

As already noted, the HB lifetime does strongly depend also on the efficiency of convection-induced mixing at the boundary
of the convective core. More
in detail, all along the HB evolutionary phase, the treatment of mixing at the boundary of the convective core,  is really a
relevant problem. In fact, as a consequence of the burning processes, He is transformed into carbon and oxygen whose associate
opacity is larger with respect that of an He rich mixture. This change in opacitive properties of the stellar matter in the core,
strongly modifies the behaviour of the radiative gradient, producing an increasing of the mass size of the convective core.

{\small
\begin{figure}
\hspace*{1.5cm}\psfig{figure=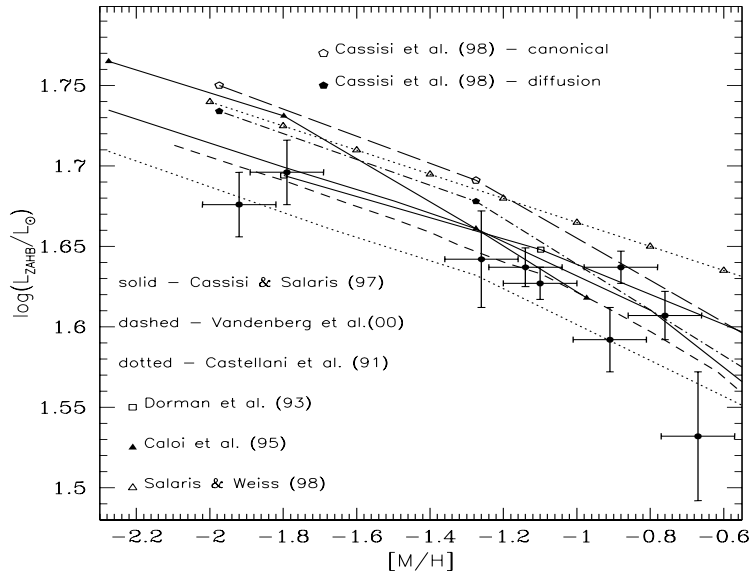,height=8truecm,width=10truecm}
\vspace*{-0.5cm}\caption{The trend of the ZAHB luminosity as a function of the global metallicity
as provided by different authors (see labels). For the sake of comparison, the semi-empirical measurements
of the ZAHB luminosity level in a selected sample of GGCs as obtained by De Santis \& Cassisi~(1999) are also shown.}
\label{fig16}
\end{figure}
}

Unfortunately, in spite of the many theoretical works published over the last three decades, the physics that 
determines the extent of this convective region is still poorly known. The theoretical calculations 
available so far leave various scenarios open. Classical models, those based on a bare Schwarzschild criterion 
(Iben \& Rood 1970), are still calculated and widely used in many studies 
(e.g., Althaus et al. 2002). However, models that include some algorithm 
to handle the discontinuity of the opacity that forms at the external border of the convective core 
as a consequence of the conversion of He into C and O should be considered as more reliable 
(Castellani, Giannone, \& Renzini 1971a, 1971b; Demarque \& Mengel 1972; Sweigart \& Gross 1976; Castellani et al. 1985; 
Dorman \& Rood 1993 and references therein). According to this mixing scheme, the change in the opacitive properties of the core 
naturally leads to the growth of the convective core (the so-called induced overshoot) and to the formation of a 
semiconvective layer outside the fully convective region. 

In an alternative approach, it is assumed that a mechanical overshoot 
takes place at the boundary of the convective region (Saslaw \& Schwarzschild~1965; Girardi et al.~2000). 
Although, the real occurrence of this phenomenon is out of debate, a quantitative estimate
of the overshoot efficiency is still a unsettled issue. 
However, detailed evolutionary computations (Straniero et al.~2003) show that a moderate efficiency of mechanical overshoot 
mimics the effect of induced
overshoot, whereas a large efficiency would produce a convective core so large to include the semiconvective region, so causing
large changes in the structural and evolutionary properties. 

The approach adopted for managing the convection-induced mixing in HB structures does affect largely the evolutionary lifetimes
as a consequence of the change in available amount of fuel, but in addition it largely affects also the C/O ratio in the CO core
at the central He exhaustion. Therefore the effects are quite similar to different assumptions about the rate for the nuclear process
${\rm {^{12}C}(\alpha,\gamma){^{16}O}}$. Therefore, there exists a sort of degeneracy between this reaction rate and the efficiency of
mixing during the core He-burning phase. The question is how we can break this degeneracy.

The answer is positive: this can be obtained by using different independent empirical constraints whose comparison with
theoretical predictions allows to disentangle the evolutionary and structural effects associated with nuclear reaction rates and
mixing processes. In this context, it is useful to remember that in these last years a large effort has been devoted to the
calibration of the R parameter (Buzzoni et al. 1983) in order to estimate the primordial He abundance of the GGC system. This
parameter is defined as the number ratio between HB and RGB stars brighter than the HB. So its theoretical calibration is strongly
affected by model predictions about the HB lifetime. The recent analysis performed by Cassisi et al. (2003) and Salaris et al.
(2004) have shown that the new generation of HB models based on the more recent evaluation of the ${\rm
{^{12}C}(\alpha,\gamma){^{16}O}}$ rate and on the semiconvective mixing scheme\footnote{These models neglect also the occurrence of
breathing pulses in the late phase of the core He-burning stage. For a detailed discussion about the reasons for which the
occurrence of this process in real stars is considered implausible we refer to Caputo et al.~(1989).} are able to provide an estimate of the initial He abundance in very good agreement with the measurements obtained through
the analysis of the Cosmic Microwave Background anisotropies and primordial nuclesynthesis models. They also predict
a value for the parameter $R_2$ (i.e. the ratio between the number of AGB stars and that of HB objects - Caputo, Castellani \& Wood~1978) 
in fair agreement with observations. It is worth noting that the $R_2$ parameter is strongly affected by the adopted
mixing approach, since the larger the mixing during the core He-burning, the less the amount of He will be available for
the subsequent AGB evolutionary phase.

In addition, the recent analysis of the non-radial pulsations of White Dwarfs (Metcalfe et al.~2000, 2001) can provide important clues
about the C/O ratio within the CO core as well as on the ratio between the CO core mass and the total WD mass. All these
empirical constraints when analyzed within a self-consistent, and updated evolutionary framework can be of extreme relevance in
order to improve our knowledge on the physical processes at work in He-burning, low-mass stars.

\section{The clump of the Asymptotic Giant Branch}

Stellar evolutionary models (Castellani, Chieffi, \& Pulone 1990) consistently predict that 
after the central He-exhaustion, the He burning rapidly moves from the core to the shell surrounding the CO core whose extension
is fixed by the mass size of the convective core during the previous HB phase. 
Thus, the beginning of the AGB is characterized by a rapid increase in luminosity. When the shell He burning 
stabilizes, a slowing down of the evolutionary rate is expected. These theoretical predictions are well confirmed by empirical evidence
in GGCs showing that the transition between the central and the shell He burning is marked by a clear gap (where few stars are observed), 
and that a well-defined clump of stars (at least in the more populous - or the well sampled - clusters) is found at 
the base of the AGB. 

From a theoretical point of view (Pulone 1992; Bono et al. 1995) it is well known that the luminosity level 
of the AGB clump is almost independent of the chemical composition, i.e. it does not depend significantly on the initial He abundance and
metallicity (see fig.~\ref{fig17}). As a consequence, it was suggested by Pulone (1992) the use of this observational feature as standard candle. 

{\small
\begin{figure}
\vspace*{-0.3cm}  
\hspace*{1.5cm}\psfig{figure=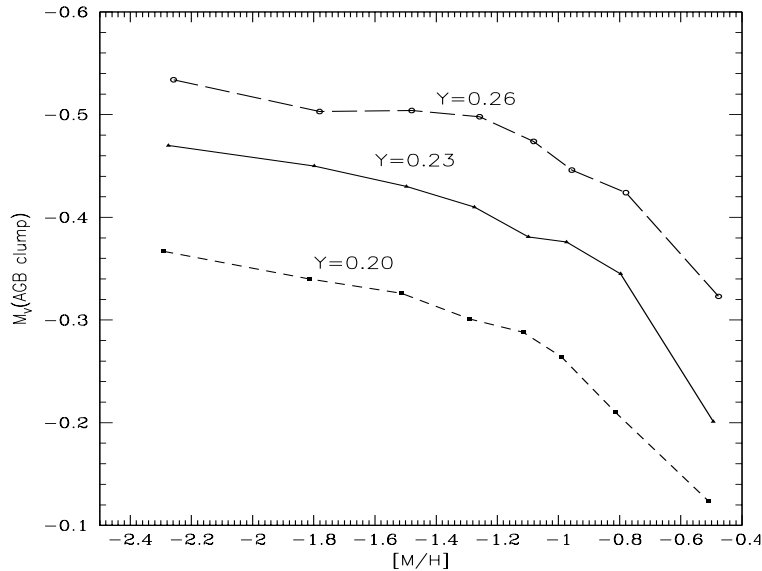,height=8truecm,width=10truecm}
\vspace*{-0.5cm}\caption{The trend of the absolute visual magnitude of the AGB clump 
as function of the global metallicity for different initial He abundances as predicted by ouw own models.}
\label{fig17}
\end{figure}
}

When checking
the reliability of theoretical predictions about the luminosity of the AGB clump by comparison with empirical evidence, in order to
overcome the problem related with the quite uncertain cluster distance scale, it is common to consider, in analogy with the RGB bump, the
brightness (for instance in the V band) difference between the AGB clump and the HB, i.e. the parameter $\Delta{V_{HB}^{AGB}}$.

A detailed analysis of the effects on the AGB clump brightness associated to the current uncertainties in the adopted physical inputs as
well as in the approach adopted for treating the mixing processes during the core He-burning phase has been performed by Cassisi et al.
(2001). They found that the $\Delta{V_{HB}^{AGB}}$ parameter is not affected at all by current uncertainty in the atomic diffusion
coefficients. 

However, this quantity is quite sensitive to the approach used for the treatment of the breathing pulses phenomenon 
at the end of the core He-burning stage. 
More in detail, we note that the empirical evidence of the AGB clump in GGCs seems to clearly rule out the occurrence of
this phenomenon in real stars. In fact, numerical simulations show that, when breathing pulses are allowed to occur, the drop in luminosity
associated to the AGB clump is almost vanishing. As a consequence, population synthesis models, based on stellar models accounting for the occurrence
of breathing pulses, do not show any evident increase in the star count in the region of the CMD where the AGB clump is really observed. 

In addition, Cassisi et al. (2001) have shown
that, at odds with what occur for the HB and AGB evolutionary lifetimes (see the discussion in section~\ref{hbtime}), 
the $\Delta{V_{HB}^{AGB}}$ parameter is not largely affected by present uncertainties in the physical inputs adopted for computing 
the stellar models.

Recently Ferraro et al. (1999) have investigated the dependence of the AGB clump brightness on the HB
morphology. The existence of a clear correlation between the brightness of the clump and the HB type, in the sense that old stellar
systems with bluer HBs are expected to show an AGB clump becoming bluer and bluer and less clumpy (and, in turn, less observable) has to
be accounted for before using this feature as a distance indicator.

\section{Final remarks}

In this paper (but see also the rich, quoted, literature), we have shown that theoretical predictions on stellar models are affected by sizeable
uncertainties, a clear proof being the occurrence of not-negligible differences between results provided by different theoretical groups. 

From the point of view of stellar models users, the best approach to be used for properly accounting for these uncertainties, is to not use evolutionary
results with an uncritical approach and, also to adopt
as many as possible independent theoretical predictions in order to have an idea of the uncertainty existing in the match between
theory and observations. It would be also worthwhile to pay attention to the improvements adopted both in the physical inputs as well as in the
physical assumptions, by people computing the evolutionary models.

On the other hand, stellar model makers should continue their effort of continuously updating their models in order to account for the
\lq{best}\rq\ physics available at any time, and consider the various empirical constraints as a benchmark of their stellar models. This
represents a fundamental step for obtaining as much as possible accurate and reliable stellar models.

In the previous sections, we have mentioned that a quite important source of uncertainty in the comparison between theory and observations derives from
the errors still affecting both theoretical and empirical color - effective temperature relations and the bolometric correction scales for the
different photometric bands. It is evident that these uncertainties do strongly hamper the possibility of a sound comparison between stellar models
and empirical evidence and, of course, make extremely problematic to assess the level of accuracy of present evolutionary scenario. In our belief, a big
effort should be devoted in the near future in order to improve the accuracy and reliability of the transformations adopted for transferring stellar
models from the H-R diagram to the various observational planes.

It has also been emphasized the huge impact of both GGCs metallicity and distance scale uncertainties on the possibility to realize a meaningful
comparison between theory and observations. Although, large improvements have been achieved in these fields, current errors are still too large for
offering the opportunity of a plain assessment of residual uncertainties in the present generation of stellar models.

These considerations make clear that a sizeable improvement in the stellar evolution framework could be achieved, in the near future, only if scientists,
working in different fields of astrophysical research, will provide their own contribution to reduce the still existing uncertainties affecting both
the theoretical framework as well as the observational scenario.

\acknowledgements           
We warmly acknowledge G. Bono, V. Castellani and A. Irwin for their pertinent suggestions and positive comments on an
early draft of this manuscript. We are also grateful to them for many enlightening discussions. We also wish to thank M. Bellazzini, 
S. Degl'Innocenti and A. Pietrinferni for providing the data shown in some figures. We warmly thank the LOC and the SOC for
organizing this interesting meeting. It is a real pleasure to thank David Valls-Gabaud for all the help provided and the pleasant
discussions. This research has made use of NASA's Astrophysics Data System Bibliographic Services.


\end{document}